\documentclass[manuscript]{aastex}
\usepackage{lineno}
\usepackage{bm}
\usepackage{color}
\usepackage{esint}
\usepackage{appendix}
\usepackage{multirow}
\usepackage{makecell}
\usepackage{threeparttable}
\usepackage{url}
\usepackage{ragged2e}
\usepackage{longtable}
\usepackage{rotating}
\usepackage{footnote}
\usepackage{diagbox}

\begin{document}
\title{Successive Coronal Mass Ejections Associated with Weak Solar Energetic Particle Events}
\author{Bin Zhuang\altaffilmark{1}, No\'{e} Lugaz\altaffilmark{1}, Tingyu Gou\altaffilmark{2,3}, and Liuguan Ding\altaffilmark{4}}
	
\affil{$^1$Institute for the Study of Earth, Ocean, and Space, University of New Hampshire, Durham, NH, USA; \url{bin.zhuang@unh.edu}, \url{noe.lugaz@unh.edu} \\
$^2$CAS Key Laboratory of Geospace Environment, Department of Geophysics and Planetary Sciences University of Science and Technology of China, Hefei 230026, China \\
$^3$CAS Center for Excellence in Comparative Planetology, Hefei 230026, People’s Republic of China \\ 
$^4$School of Physics and Optoelectronic Engineering, Nanjing University of Information Science and Technology, Nanjing 210044, China} 

\begin{abstract}
The scenario of twin coronal mass ejections (CMEs), i.e., a fast and wide primary CME (priCME) preceded by previous CMEs (preCMEs), has been found to be favorable to a more efficient particle acceleration in large solar energetic particle (SEP) events. Here, we study 19 events during 2007--2014 associated with twin-CME eruptions but without large SEP observations at L1 point. We combine remote-sensing and in situ observations from multiple spacecraft to investigate the role of magnetic connectivity in SEP detection and the CME information in 3-dimensional (3D) space.
We study one-on-one correlations of the priCME 3D speed, flare intensity, suprathermal backgrounds, and height of CME-CME interaction with the SEP intensity. Among these, the priCME speed is found to correlate with the SEP peak intensity at the highest level. We use the projection correlation method to analyze the correlations between combinations of these multiple independent factors and the SEP peak intensity. We find that the only combination of two or more parameters that has higher correlation with the SEP peak intensity than the CME speed is the CME speed combined with the propagation direction. This further supports the dominant role of the priCME in controlling the SEP enhancements, and emphasizes the consideration of the latitudinal effect. Overall, the magnetic connectivity in longitude as well as latitude and the relatively lower priCME speed may explain the existence of the twin-CME SEP-poor events. The role of the barrier effect of preCME(s) is discussed for an event on 2013 October 28. 
\end{abstract}


\justifying
\section{Introduction}
A two-class paradigm for impulsive and gradual solar energetic particle (SEP) events was generally accepted by the end of 1990's \citep[e.g.,][]{1999SSRv...90..413R,2013SSRv..175...53R,2016LRSP...13....3D}. Gradual and large (flux of ions with energy above 10 MeV greater than 10 pfu, $1 \ \rm{pfu}=1 \ \rm{proton} \ \rm{cm}^{-2} \ \rm{s}^{-1} \ \rm{sr}^{-1}$) SEP events are believed to be primarily generated by shocks driven by fast and wide coronal mass ejections (CMEs), while impulsive events are thought to be generated by magnetic reconnection in solar flares. Understanding the acceleration mechanism of SEPs is one of the outstanding problems in space science research, and most researchers accepted the notion that large gradual SEP events result from the diffusive shock acceleration \citep{1983JGR....88.6109L}. To explain the particle acceleration in the much stronger ground level enhancement events (GLEs) compared to the normal gradual SEP events, a twin-CME scenario was proposed by \citet{2012SSRv..171..141L}. In this scenario, the CME which accelerates SEPs (called primary CME, priCME) is associated with a preceding CME (preCME). The preCME can contribute to a much more efficient particle acceleration by providing an excess of seed populations and enhanced turbulence level for the shock driven by the priCME. This initial scenario was further investigated in the last decade. \citet{2013ApJ...763...30D} statistically analyzed the large SEP events in solar cycle (SC) 23, and they found that the events associated with twin-CME eruptions have on average a higher peak intensity ($I_p$) than those associated with single-CME eruption. The case studies on the effect of CME-CME interaction for the events on 2012 May 17 and 2013 May 22 were provided by \citet{2013ApJ...763..114S} and \citet{2014ApJ...793L..35D}, respectively, which further supports the twin-CME scenario in SEP enhancements. Besides, the effect of preconditioning by preCMEs had also been studied previously by \citet{2002ApJ...572L.103G,2003GeoRL..30.8015G,2004JGRA..10912105G}, and \citet{2012AIPC.1436..247G} found that in SC23 SEP events associated with preCMEs have relatively higher $I_p$ than those not associated with preCMEs.

Later on, the role of CME-CME interaction to intensity SEP enhancements was challenged by \citet{2014ApJ...784...47K} They found that the timings of the preCMEs relative to the priCMEs, and the widths, speeds, and numbers of the preCMEs do not correlate with the SEP $I_p$; but the 2-MeV proton backgrounds correlate with the SEP $I_p$ and the fraction of CMEs with preCMEs. They suggested that it is not the CME-CME interaction, but the general increase of background seed particles and more frequent CMEs during times of higher solar activity that result in the higher SEP $I_p$. \citet{2020ApJ...901...45Z} further studied the acceleration and release of SEPs associated with different types of CME-CME interaction. They found that a meaningful CME-CME interaction is not necessary for the efficient acceleration and release of SEPs; however, events for which the priCME and preCME are close to each other at the particle release time tend to be associated with a relatively higher $I_p$.

In the complex conditions (with or without twin-CME eruption) of SEP acceleration and transport, many factors could contribute to controlling the SEP peak intensity. Based on the general CME-SEP relationship, a large number of statistical studies have shown the correlations between CME properties (e.g., speed, angular width, or kinetic energy) vs. SEP $I_p$ \citep[e.g.,][]{2004JGRA..10912105G,2013ApJ...763...30D,2005JGRA..11012S01K,2014ApJ...784...47K,2014SoPh..289.3059R,2015SoPh..290..819T,2019JGRA..124.6384X,2020ApJ...900...75K,2020ApJ...901...45Z}. Moreover, the relationship between the shock properties and SEPs was also analyzed \citep{2015ApJ...799..167K,2019ApJ...876...80K}, as the shock strength is determined not only by the shock speed but also by the background (plasma density and magnetic fields). \citet{2019ApJ...876...80K} found a high correlation between the SEP $I_p$ and the strength of the shock (quantified, e.g., by the Mach number) in the corona.

The role of the flare acceleration in large SEP events has been widely analyzed \citep[e.g.,][]{2010JGRA..115.8101C,2013SoPh..282..579M,2014JGRA..119.9456P,2015SoPh..290..841D,2015SoPh..290..819T,2016ApJ...833L...8T,2020ApJ...900...75K}. The soft X-ray (SXR) peak flux associated with flares was found to be statistically correlated with the SEP peak  \citep{2010JGRA..115.8101C,2013SoPh..282..579M,2015SoPh..290..841D,2016ApJ...833L...8T}. This suggests that flares are an important contributor in accelerating SEPs. However, the contribution of flares in the principal source of high-energy protons is still under discussion. \citet{2014JGRA..119.9456P} studied the occurrences of solar proton events with CMEs and flares, and found that the CME parameters, not the flare SXR peak flux, act as the most important parameters in understanding SEP peak intensities. \citet{2015SoPh..290..819T} used the partial correlation method to remove the correlation effects between solar parameters themselves, and found that the CME speed and the SXR fluence are the only statistically significant contributions for the SEP peak intensity.

The role of the seed populations is expected to be an important parameter in the diffusive shock acceleration both for a single CME or in the twin-CME scenario. Several studies have suggested that the suprathermal particles measured in situ near 1 AU represent a major portion or a good proxy for of seed ion populations accelerated near the Sun \citep{2005AIPC..781..219M,2006ApJ...649..470D,2012AIPC.1436..130G}. \citet{2014ApJ...784...47K} and \citet{2014ApJ...791....4K} found relatively high correlation between the 2-MeV (suprathermal) proton intensities before an SEP event and the resulting peak intensities for protons $>20$ MeV during SEP events.  \citet{2019ApJ...872...89K} further analyzed the correlations by using suprathermal particle (H and He) observations and considered different types of solar transients from eastern and western solar hemisphere. They confirmed that tracking suprathermal intensities at 1 AU can
be useful in forecasting SEP $I_p$. \citet{2015ApJ...812..171D} studied the seed populations in the large SEP events associated with the twin-CME eruptions, and found that the abundance of iron to oxygen ratio is at a relatively higher level compared to that associated with single CME eruption. This indicates that iron-rich flare materials may be from preceding flares, and may play an important role in the twin-CME associated SEP events.

In addition to the conditions described above, magnetic connectivity between the observing spacecraft and the SEP source region is one of the key factors in determining the in situ observed SEP level. The launch of the twin Solar TErrestrial RElations Observatory \citep[STEREO, STA and STB hereafter;][]{2008SSRv..136....5K} extended the community's ability to study the longitudinal distribution of SEPs \citep[e.g.,][]{2012ApJ...752...44R,2013ApJ...767...41L,2014A&A...567A..27D,2014SoPh..289.3059R,2019JGRA..124.6384X}. A Gaussian expression with a constant distribution width is typically applied to the observations to parametrize the longitudinal distribution of SEP intensities. \citep[see][]{2013ApJ...767...41L,2014SoPh..289.3059R}. Recently, \citet{2019JGRA..124.6384X} found that the distribution width is dependent on the connection between the observer and the source region with an east-west asymmetry. Case studies on the longitudinal properties of widespread SEP events, e.g., \citet{2016ApJ...821...31K}, \citet{2016ApJ...819...72L} and \citet{Palmerio_2021}, have further highlighted the importance of using multiple spacecraft in SEP investigations.

While the positive CME-SEP relationship has been widely studied, the opposite has not been investigated in as much depth, i.e., why a fast and wide CME does not cause a strong SEP event, or whether or not a slow or narrow CME can lead to the SEP event with higher $I_p$. \citet{2020ApJ...889...92L} focused on the fast and wide CMEs without observed $>$25 MeV protons, and they proposed some physical reasons for the lack of SEP observations: (1) the driving CMEs are relatively narrower and slower than those associated with large SEP events, (2) the preceding solar eruptions may not release seed particles, (3) the primary eruptions may not release SEPs which contribute directly to the prompt component of the event, and (4) the magnetic connectivity between the source regions and spacecraft is poor. As for the second point, \citet{2019SoPh..294..134K} confirmed that no narrow CMEs can be associated with large SEP events. They explained this in terms of a distinction between bow shocks around narrows CMEs and piston-driven shocks around wide CMEs. 

In the twin-CME scenario, the particles are expected to be accelerated more efficiently. Here, we focus on whether or not there exist twin-CME events not associated with SEP observations even when the priCME is fast and wide. If so, what are the reasons for this phenomenon? These two questions are the motivation of this work. In addition, \citet{2019RAA....19....5D} analyzed the relationship between large SEP events and twin-CME events in SC24, but they did not limit the speed of the preCMEs and priCMEs, and thus slow CMEs were also considered. They found that, for their samples, large SEP events were all generated by CMEs associated with the presence of enhanced type II radio bursts. The enhancement of type II radio bursts is supposed to be an indicator for the occurrence of CME-CME interaction \citep{2001ApJ...548L..91G,2014ApJ...793L..35D,2017SoPh..292...64L}.

The paper is arranged as follows. In Section \ref{sec_ins}, we describe the instruments used and the intercalibration between different instruments. In Section \ref{evanaly}, we first describe the identification of the twin-CME SEP-poor events, and then show one-on-one correlations between the potential controlling factors (including the CME speed, flare intensity, suprathermal backgrounds and the CME-CME interaction) and the SEP peak intensity. In Section \ref{sec_dis}, we perform the projection correlation method to analyze the relationship between the combination of these multiple controlling factors and one dependent SEP peak intensity, and we provide further discussions including the analysis for the event on 2013 October 28. Section \ref{summary} summaries our findings and concludes.

\section{Instruments and Data Preparation}\label{sec_ins}
In this paper, we use the in situ instruments of the Energetic Particle Sensor (EPS) suite on board Geostationary Operational Environmental Satellites (GOES), the Energetic and Relativistic Nuclei and Electron instrument \citep[ERNE with the Low Energy Detector (LED) and the High Eneryg Detector (HED);][]{1995SoPh..162..505T} on board the SOlar and Heliospheric Observatory \citep[SOHO;][]{domingo95}, and the Low Energy Telescope \citep[LET,][]{2008SSRv..136..285M} and the High Energy Telescope \citep[HET,][]{2008SSRv..136..391V} on board STEREO for the SEP observations. The Solar Electron and Proton Telescope \citep[SEPT;][]{2008SSRv..136..363M} on board STEREO, and the Electron, Proton, and Alpha Monitor \citep[EPAM;][]{1998SSRv...86..541G} on board the Advanced Composition Explorer (ACE) are used for the observations of suprathermal backgrounds. The Large Angle and Spectrometric Coronagraph \citep[LASCO;][]{1995SoPh..162..357B} on board SOHO and the coronagraphs \citep[COR1 and COR2;][]{2008SSRv..136...67H} on board STEREO are used for the CME observations. The Extreme Ultraviolet Imager \citep[EUVI;][]{2008SSRv..136...67H} on board STEREO and the Atmospheric Imaging Assembly \citep[AIA;][]{2012SoPh..275...17L} on board Solar Dynamics Observatory \citep[SDO;][]{2012SoPh..275....3P} are used for the observations of the eruptive signatures on the solar surface. The catalogs of CMEs \citep[\url{https://cdaw.gsfc.nasa.gov/CME_list/index.html};][]{2004JGRA..109.7105Y} and large SEP events \citep[\url{https://cdaw.gsfc.nasa.gov/CME_list/sepe/};][]{2009EM&P..104..295G} from the Coordinated Data Analysis Workshop (CDAW) Data Center are also used.

\begin{figure}[!hbt]
	\centering
	\includegraphics[width=0.97\textwidth]{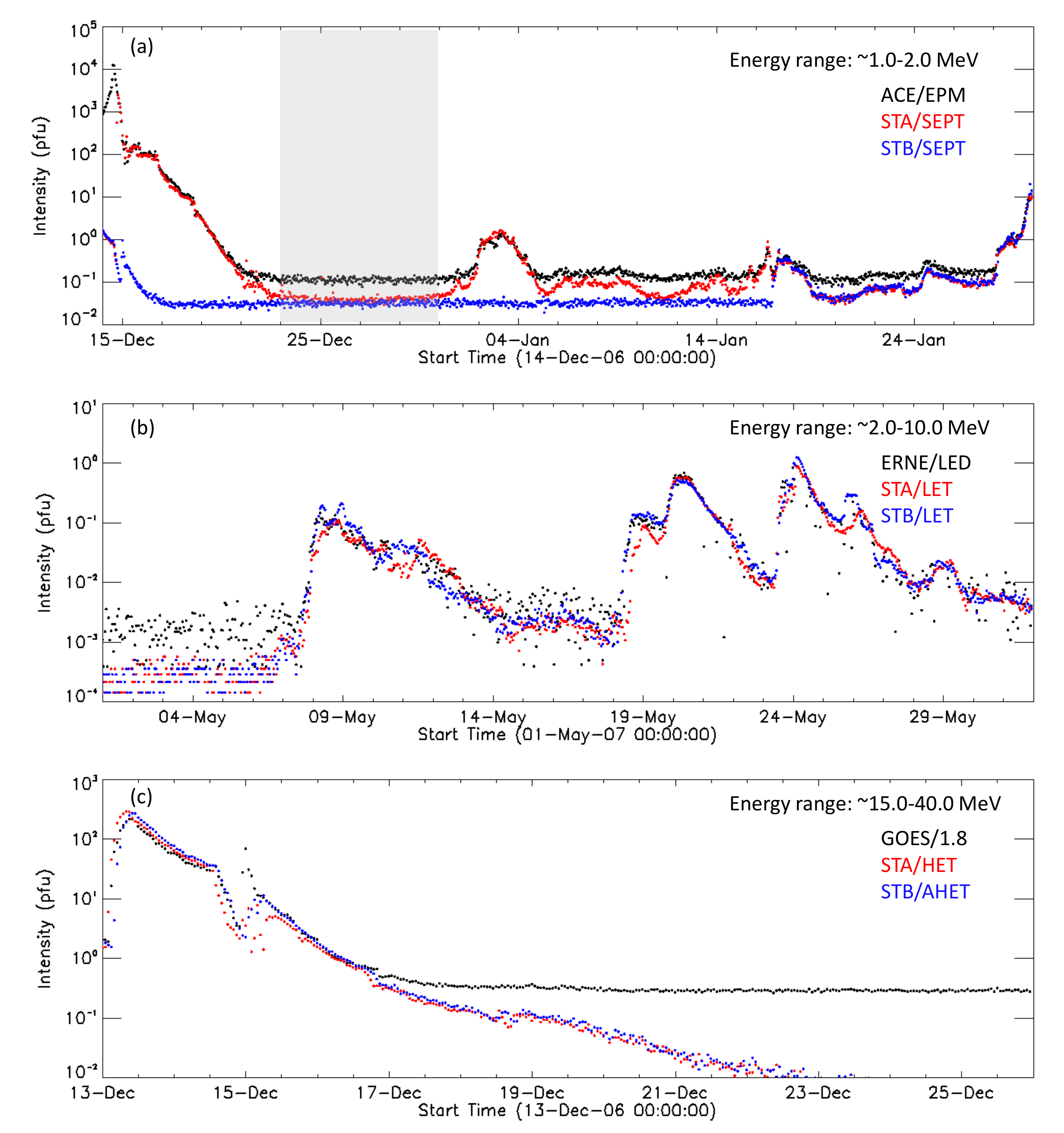}
	\caption{\small (a) Proton intensities observed by ACE/EPAM and STEREO/SEPT from 2006 December 14 to 2007 January 30. (b) Proton intensities observed by ERNE/LED and STEREO/LET from 2007 May 1 to 2007 May 31. (c) Proton intensities observed by GOES/EPS and STEREO/HET from 2006 December 13 to 2006 December 26. } 
	\label{sc_intercali}
\end{figure}

As for the in situ energetic proton observations, we consider three energy ranges, including (1) 1.06--1.91 MeV at ACE/EPAM vs. 1.11--1.99 MeV at STEREO/SEPT, (2) 2.0--10.0 MeV at ERNE/LED vs. 2.2-10.0 MeV at STEREO/LET, and (3) 15.0--40.0 MeV at GOES/EPS vs. 14.9--40.5 MeV at STEREO/HET. Before obtaining the SEP intensities, the intercalibration between different instruments are needed. Figure \ref{sc_intercali} shows the comparisons of the hourly intensities between different energy ranges and different spacecraft.
(a) shows the profiles of the proton intensities observed by ACE/EPAM and STEREO/SEPT in $\sim$1--2 MeV. The intensity enhancements at three spacecraft match well with each other since 2007 January; but the background level during quiet time (e.g., shown by the shadow) at EPAM is higher than that at SEPT. Therefore, we calibrate SEPT data by multiplying a factor of 2.5 for intensities lower than 0.1 pfu. (b) shows the profiles of the proton intensities observed by ERNE/LED and STEREO/LET in $\sim$2--10 MeV in 2007 May (there is no ERNE/LED data before 2007 May). Although the angular separations between the three spacecraft are not trivial at this time ($\sim 2$--$8^\circ$ between each of the STEREO and ERNE), the global magnitudes and trends still match quite well with each other. Thus, no additional intercalibration is performed here. (c) shows the profiles of the proton intensities observed by GOES/EPS and STEREO/HET in $\sim$15--40 MeV. It is found that the corrected GOES profile (divided by 1.8) matches well with the HET profiles when the intensity is higher than $\sim 0.3$ pfu due to the different instrument background level; the GOES intensity lower than 0.3 pfu is not used.

\section{Twin-CME SEP Events}\label{evanaly}
\subsection{Identification of Twin-CME SEP-Poor Events}\label{evinfo}
We start by identifying the twin-CME SEP-poor events. During identification, SEP-poor events are associated with peak intensity $<10$ pfu for protons $>10$ MeV. The identification procedures are described as follows: (1) find potential priCMEs with speed greater than 800 ${\rm{km \, s ^ {-1}}}$ and angular width larger than $60^\circ$ based on CDAW/CME catalog during 2007--2014; (2) find potential preCME(s) associated with every priCME within 12 hours based on the LASCO/C2 first appearance time, and the CDAW speed and angular width of a preCME are required to be greater than 300 km/s and $20^\circ$, respectively; (3) combine the observations from STA and STB, and the graduated cylindrical shell (GCS) model \citep{2006ApJ...652..763T,2009SoPh..256..111T} to check whether the priCME and preCME(s) have spatial overlap in 3D space, while the CME 3D parameters derived by the GCS model for the preCME and priCME also need to meet the criteria of the speeds and angular widths above; and (4) eliminate the L1-point-observed large SEP events with peak intensity $>10$ pfu for protons $>10$ MeV (on CDAW/SEP website) which have been analyzed by \citet{2020ApJ...901...45Z}. The GCS model assumes a CME with a flux-rope structure, and has six free parameters for the model reconstruction: the latitude ($\theta$) and longitude ($\phi$) of the CME propagation direction, the height of the CME leading edge ($h$), the aspect ratio ($\kappa$), the tilt angle ($\gamma$) with respect to the equator, and the half angular width ($\delta$) between the two flux rope legs. The CME speed is derived by a linear fit to the $h$-time measurements, and all the parameters are set to be constant during the CME propagation. One can refer to \citet{2020ApJ...901...45Z} for the details of the usage of the GCS model. The event identification procedure is the same as the one applied for twin-CME SEP-rich events in \citet{2020ApJ...901...45Z} except that we do not have a threshold for the flux of energetic protons. 

We identified 19 events, and Table \ref{sep_ip} lists the related information, including (1) $I_p$ in 2--10 MeV and 15--40 MeV at three spacecraft if measured, and (2) the SEP onset and peak times observed by the spacecraft with the highest $I_p$ in 15--40 MeV. The intensities are background-corrected, the peak associated with the passage of the interplanetary shock is not considered, and no intercalibration factor has been applied in the intensities listed in the tables. Due to the weak solar activity during solar minimum, the twin-CME events associated with fast and wide priCME started to occur in 2011. The L1-point large SEP events associated with twin-CME eruptions in SC24 in \citet{2020ApJ...901...45Z} are also considered in this work. Table \ref{large_sep} lists the corresponding information, similar to Table \ref{sep_ip}. As compared to the work of \citet{2020ApJ...901...45Z}, we add the $I_p$ at the STEREO spacecraft. Table \ref{cmeinfo} has the information about the preCMEs and priCMEs for the events in Table \ref{sep_ip}, including the LASCO/C2 first appearance time, the 2D parameters from the CDAW catalog of the central position angle, angular width and speed of the CMEs, and the 3D parameters by the GCS model of the propagation direction, speed, tilt angle and face-on angular width $w_f=2\delta+w_e$ and edge-on width $w_e=2\sin ^{-1} (\kappa)$. The related CME information for the events in Table \ref{large_sep} can be found in \citet{2020ApJ...901...45Z}.

\footnotesize
\begin{longtable}{|c|c|c|c|c|c|c|c|c|}
	\caption{Twin-CME (L1-point) SEP-poor events during 2007--2014} 
	\label{sep_ip} \\ \hline
	\multicolumn{2}{|c}{SEP Event} & \multicolumn{2}{|c}{$I_p$(B)} & \multicolumn{2}{|c|}{$I_p$(E)} & \multicolumn{2}{|c|}{$I_p$(A)} & $I_0$ \\ \hline
	Onset & Peak & 2--10 & 15--40 &  2--10 & 15--40 & 2--10 & 15--40 & 15--40 \\ \hline
	2011-05-29 22:00 & 05-30 08:25 & 782.38 & 3.87 & - & - & - & - & 23.38 \\ \hline
	2011-09-24 13:00 & 09-25 06:20 & 2102.93 & 149.74 & - & - & - & - & 394.04 \\ \hline
	2011-10-14 15:00 & 10-15 09:50 & - & 0.06 & - & - & 18.49 & 0.14 & 6.26 \\ \hline
	2011-10-22 12:00 & 10-23 08:00 & 0.92 & 0.12 & 413.95 & 1.82 & 25.90 & 0.28 & 2.77 \\ \hline
	2011-11-09 14:30 & 11-09 19:00 & 24.91 & 0.33 & - & - & - & - & 0.95 \\ \hline
	2012-01-16 10:00 & 01-17 08:00 & 98.93 & 0.53 & - & - & - & - & 0.94 \\ \hline
	2012-03-04 15:00 & 03-05 02:30 & 1245.60 & 66.85 & - & - & 0.13 & 0.06 & 81.35 \\ \hline
	2012-03-05 08:00 & 03-05 12:00 & 479.51 & 20.92 & 44.38 & 1.85 & - & 0.18 & 22.12 \\ \hline
	2012-04-18 17:00 & 04-18 23:00 & - & 0.04 & - & - & 1.02 & 0.07 & 0.59 \\ \hline
	2012-08-21 21:00 & 08-22 06:00 & 0.54 & 0.09 & - & - & - & - & 6.44 \\ \hline
	2012-09-20 16:00 & 09-21 06:00 & 56.22 & 6.42 & - & - & 868.42 & 142.57 & 390.88 \\ \hline
	2013-07-03 08:00 & 07-03 11:20 & - & 0.23 & - & - & - & - & 0.28 \\ \hline
	2013-07-22 08:00 & 07-22 10:20 & 2.03 & 0.15 & - & - & 85.74 & 1.97 & 9.59 \\ \hline
	2013-10-25 18:00 & 10-25 20:40 & 129.68 & 23.90 & - & - & 0.11 & 0.08 & 25.27 \\ \hline
	2013-10-28 05:30 & 10-28 14:00 & - & - & 79.90 & 1.64 & 0.20 & 0.19 & 0.91 \\ \hline
	2014-01-30 21:00 & 01-31 09:00 & 9.84 & 0.23 & - & - & - & 0.11 & 1.23 \\ \hline
	2014-05-07 18:00 & 05-07 21:15 & - & - & 6.78 & 0.58 & - & - & 1.46 \\ \hline
	2014-06-06 15:00 & 06-06 17:20 & - & - & - & - & 16.83 & 0.58 & 1.24 \\ \hline
	2014-06-10 16:00 & 06-10 21:00 & 109.76 & 4.78 & - & - & 33.96 & 0.41 & 5.24 \\ \hline
\end{longtable}
\begin{tablenotes}	
	\item[1] [1] The table lists the information of the L1-point SEP-poor events during 2007--2014, including the SEP onset and peak time based on the observations at one spacecraft with highest $I_p$ with $15<E<40$ MeV, $I_p$ (unit: pfu) at 2--10 and 15--40 MeV at three spacecraft, and the estimated peak intensities ($I_0$) which has the best magnetic connectivity between the source region and the observing spacecraft. All the $I_p$ are background-corrected. B: STEREO-B, E: Earth (L1 point), and A: STEREO-A.
\end{tablenotes}

\begin{longtable}{|c|c|c|c|c|c|c|c|c|}
	\caption{Twin-CME (L1-point) large SEP events in solar cycle 24 (adapted from \citet{2020ApJ...901...45Z})} 
	\label{large_sep} \\ \hline
	\multicolumn{2}{|c}{SEP Event} & \multicolumn{2}{|c|}{$I_p$(B)} & \multicolumn{2}{|c}{$I_p$(E)} & \multicolumn{2}{|c|}{$I_p$(A)} & $I_0$ \\ \hline
	Onset & Peak & 2--10 & 15--40 &  2--10 & 15--40 & 2--10 & 15--40 & 15--40 \\ \hline
	2011-08-09 08:30 & 08-09 09:05 & - & - & 362.69 & 21.27 & - & 0.20 & 38.29 \\ \hline	
	2011-11-26 08:30 & 11-26 23:20 & 3.78 & 0.18 & 1944.95 & 28.43 & 496.20 & 2.01 & 53.72 \\ \hline
	2012-01-23 05:00 & 01-24 15:30 & 225.50 & 38.41 & 4951.90 & 2858.58 & 682.76 & 23.28 & 9120.00 \\ \hline
	2012-03-07 03:00 & 03-08 03:00 & 8943.32 & 1483.22 & 4874.07 & 1365.61 & 12.54 & 3.05 & 16243.30 \\ \hline
	2012-05-17 03:00 & 05-17 07:00 & 0.25 & 0.07 & 831.10 & 191.82 & 812.74 & 0.53 & 619.11 \\ \hline
	2012-07-08 18:30 & 07-09 04:30 & - & - & - & 14.40 & - & 4.80 & 28.20 \\ \hline
	2013-05-15 07:30 & 05-15 17:50 & - & - & 960.00 & 16.46 & - & - & 629.89 \\ \hline
	2013-05-22 14:20 & 05-22 16:50 & - & - & 3308.27 & 1100.24 & 67.15 & 3.63 & 962.00 \\ \hline
	2014-01-06 08:15 & 01-06 16:00 & 0.31 & 0.20 & - & 17.84 & - & - & 100.54 \\ \hline
	2014-04-18 14:00 & 04-19 03:00 & - & - & 3930.38 & 36.08 & - & - & 115.11 \\ \hline
	2017-09-04 22:30 & 09-05 17:00 & - & - & 3330.83 & 79.25 & 0.59 & 0.14 & 665.73 \\ \hline
\end{longtable}	
\begin{tablenotes}	
	\item[1] [1] See the parameter information in Table \ref{sep_ip}.
\end{tablenotes}
\normalsize

\scriptsize
\begin{longtable}{|c|c|c|c|c|c|c|c|c|}
	\caption{CME information of the twin-CME SEP-poor events}
	\label{cmeinfo} \\ \hline
	SEP & \multicolumn{8}{|c|}{CME} \\ \hline
	Onset & 1st App. & $cpa$ & $v$ & $w$ & ($\phi, \ \theta$) & $v$ & $\gamma$ & ($w_f, \ w_e$)  \\ \hline 
	2011-05-29 22:00 & \makecell{pre 10:36 \\ pri 21:24} & \makecell{119 \\ 107} & \makecell{646 \\ 1407} & \makecell{$234^\circ$ \\ $>186^\circ$} & \makecell{($-56^\circ, \ -18^\circ$) \\ ($-69^\circ, \ -13^\circ$)} & \makecell{803 \\ 1715} & \makecell{$74^\circ$ \\ $-52^\circ$} & \makecell{($97^\circ, \ 39^\circ$) \\ ($107^\circ, \ 45^\circ$)} \\ \hline
	
	2011-09-24 13:00 & \makecell{pre 09:48 \\ pri 12:48} & \makecell{92 \\ Halo} & \makecell{1936 \\ 1915} & \makecell{$145^\circ$ \\ $360^\circ$} & \makecell{($-56^\circ, \ -2^\circ$) \\ ($-55^\circ, \ 9^\circ$)} & \makecell{1045 \\ 2104} & \makecell{$90^\circ$ \\ $90^\circ$} & \makecell{($60^\circ, \ 53^\circ$) \\ ($148^\circ, \ 53^\circ$)} \\ \hline
	
	2011-10-14 15:00 & \makecell{pre 09:12 \\ pri 12:24} & \makecell{40 \\ 32} & \makecell{454 \\ 814} & \makecell{$208^\circ$ \\ $241^\circ$} & \makecell{($233^\circ, \ 23^\circ$) \\ ($224^\circ, \ 0^\circ$)} & \makecell{470 \\ 988} & \makecell{$-90^\circ$ \\ $-59^\circ$} & \makecell{($84^\circ, \ 44^\circ$) \\ ($117^\circ, \ 50^\circ$)} \\ \hline
	
	2011-10-22 12:00 & \makecell{pre 01:25 \\ pri 10:24} & \makecell{Halo \\ Halo} & \makecell{593 \\ 1005} & \makecell{$360^\circ$ \\ $360^\circ$} & \makecell{($-295^\circ, \ 52^\circ$) \\ ($-298^\circ, \ 47^\circ$)} & \makecell{612 \\ 1130} & \makecell{$-45^\circ$ \\ $65^\circ$} & \makecell{($90^\circ, \ 40^\circ$) \\ ($150^\circ, \ 55^\circ$)} \\ \hline
	
	2011-11-09 14:30 & \makecell{pre 08:36 \\ pri 13:36} & \makecell{132 \\ Halo} & \makecell{496 \\ 907} & \makecell{$147^\circ$ \\ $360^\circ$} & \makecell{($-49^\circ, \ -32^\circ$) \\ ($-30^\circ, \ 16^\circ$)} & \makecell{568 \\ 1377} & \makecell{$13^\circ$ \\ $67^\circ$} & \makecell{($112^\circ, \ 49^\circ$) \\ ($126^\circ, \ 47^\circ$)} \\ \hline
	
	2012-01-16 10:00 & \makecell{pre 00:48 \\ pri 03:12} & \makecell{78 \\ Halo} & \makecell{708 \\ 1060} & \makecell{$161^\circ$ \\ $360^\circ$} & \makecell{($-95^\circ, \ 2^\circ$) \\ ($-61^\circ, \ 27^\circ$)} & \makecell{813 \\ 981} & \makecell{$-23^\circ$ \\ $90^\circ$} & \makecell{($57^\circ, \ 36^\circ$) \\ ($130^\circ, \ 53^\circ$)} \\ \hline
	
	2012-03-04 15:00 & \makecell{pre 05:00 \\ pri 11:00} & \makecell{47 \\ Halo} & \makecell{584 \\ 1306} & \makecell{$160^\circ$ \\ $360^\circ$} & \makecell{($274^\circ, \ 36^\circ$) \\ ($298^\circ, \ 20^\circ$)} & \makecell{470 \\ 1186} & \makecell{$9^\circ$ \\ $41^\circ$} & \makecell{($109^\circ, \ 44^\circ$) \\ ($162^\circ, \ 47^\circ$)} \\ \hline
	
	2012-03-05 08:00 & \makecell{pre 03:12 \\ pri 04:00} & \makecell{29 \\ Halo} & \makecell{594 \\ 1531} & \makecell{$92^\circ$ \\ $360^\circ$} & \makecell{($-51^\circ, \ 43^\circ$) \\ ($-53^\circ, \ 34^\circ$)} & \makecell{745 \\ 1327} & \makecell{$4^\circ$ \\ $90^\circ$} & \makecell{($85^\circ, \ 42^\circ$) \\ ($159^\circ, \ 53^\circ$)} \\ \hline
	
	2012-04-18 17:00 & \makecell{pre 09:24 \\ pri 15:12} & \makecell{64 \\ 64} & \makecell{448 \\ 840} & \makecell{$215^\circ$ \\ $184^\circ$} & \makecell{($220^\circ, \ 16^\circ$) \\ ($225^\circ, \ 16^\circ$)} & \makecell{687 \\ 1074} & \makecell{$61^\circ$ \\ $61^\circ$} & \makecell{($87^\circ, \ 41^\circ$) \\ ($101^\circ, \ 43^\circ$)} \\ \hline
	
	2012-08-21 21:00 & \makecell{pre 14:12 \\ pri 20:24} & \makecell{Halo \\ Halo} & \makecell{575 \\ 1024} & \makecell{$360^\circ$ \\ $360^\circ$} & \makecell{($-148^\circ, \ 2^\circ$) \\ ($-150^\circ, \ 2^\circ$)} & \makecell{957 \\ 1251} & \makecell{$-14^\circ$ \\ $14^\circ$} & \makecell{($68^\circ, \ 33^\circ$) \\ ($116^\circ, \ 52^\circ$)} \\ \hline
	
	2012-09-20 15:00 & \makecell{pre 05:48 \\ pri 15:12} & \makecell{Halo \\ Halo} & \makecell{663 \\ 1202} & \makecell{$360^\circ$ \\ $360^\circ$} & \makecell{($-155^\circ, \ -27^\circ$) \\ ($-150^\circ, \ -27^\circ$)} & \makecell{685 \\ 2387} & \makecell{$45^\circ$ \\ $45^\circ$} & \makecell{($160^\circ, \ 60^\circ$) \\ ($172^\circ, \ 67^\circ$)} \\ \hline
	
	2013-07-03 08:00 & \makecell{pre 00:36 \\ pri 07:24} & \makecell{126 \\ 137} & \makecell{592 \\ 807} & \makecell{$171^\circ$ \\ $>267^\circ$} & \makecell{($235^\circ, \ -23^\circ$) \\ ($236^\circ, \ -23^\circ$)} & \makecell{550 \\ 904} & \makecell{$46^\circ$ \\ $46^\circ$} & \makecell{($130^\circ, \ 40^\circ$) \\ ($130^\circ, \ 46^\circ$)} \\ \hline
	
	2013-07-22 08:00 & \makecell{pre 22:00$^{-1}$ \\ pri 06:24} & \makecell{336 \\ Halo} & \makecell{361 \\ 1202} & \makecell{$146^\circ$ \\ $360^\circ$} & \makecell{($160^\circ, \ 45^\circ$) \\ ($152^\circ, \ 20^\circ$)} & \makecell{437 \\ 1000} & \makecell{$32^\circ$ \\ $63^\circ$} & \makecell{($86^\circ, \ 43^\circ$) \\ ($163^\circ, \ 57^\circ$)} \\ \hline
	
	2013-10-25 18:00 & \makecell{pre 08:12 \\ pri 15:12} & \makecell{Halo \\ Halo} & \makecell{587 \\ 1081} & \makecell{$360^\circ$ \\ $360^\circ$} & \makecell{($-68^\circ, \ -9^\circ$) \\ ($-75^\circ, \ 5^\circ$)} & \makecell{557 \\ 1157} & \makecell{$83^\circ$ \\ $70^\circ$} & \makecell{($145^\circ, \ 49^\circ$) \\ ($171^\circ, \ 55^\circ$)} \\ \hline
	
	2013-10-28 05:30 & \makecell{pre 02:24 \\ pre 02:24 \\ pri 04:48} & \makecell{Halo \\ Halo \\ 315} & \makecell{695 \\ 695 \\ 1201} & \makecell{$360^\circ$ \\ $360^\circ$ \\ $156^\circ$} & \makecell{($-273^\circ, \ -7^\circ$) \\ ($-291^\circ, \ 25^\circ$) \\ ($-290^\circ, \ 23^\circ$)} & \makecell{563 \\ 554 \\ 1249} & \makecell{$58^\circ$ \\ $-4^\circ$ \\ $54^\circ$} & \makecell{($117^\circ, \ 47^\circ$) \\ ($92^\circ, \ 45^\circ$) \\ ($107^\circ, \ 47^\circ$)} \\ \hline
	
	2014-01-30 21:00 & \makecell{pre 15:48 \\ pri 16:24} & \makecell{109 \\ Halo} & \makecell{780 \\ 1087} & \makecell{$62^\circ$ \\ $360^\circ$} & \makecell{($-67^\circ, \ -20^\circ$) \\ ($-67^\circ, \ -36^\circ$)} & \makecell{401 \\ 1086} & \makecell{$-83^\circ$ \\ $-32^\circ$} & \makecell{($112^\circ, \ 45^\circ$) \\ ($132^\circ, \ 49^\circ$)} \\ \hline
	
	2014-05-07 18:00 & \makecell{pre 11:48 \\ pri 16:24} & \makecell{283 \\ Halo} & \makecell{811 \\ 923} & \makecell{$121^\circ$ \\ $360^\circ$} & \makecell{($-260^\circ, \ 14^\circ$) \\ ($-243^\circ, \ -22^\circ$)} & \makecell{545 \\ 919} & \makecell{$4^\circ$ \\ $-49^\circ$} & \makecell{($52^\circ, \ 33^\circ$) \\ ($136^\circ, \ 57^\circ$)} \\ \hline
	
	2014-06-06 15:00 & \makecell{pre 12:48 \\ pri 13:48} & \makecell{82 \\ Halo} & \makecell{704 \\ 1200} & \makecell{$143^\circ$ \\ $360^\circ$} & \makecell{($-118^\circ, \ 5^\circ$) \\ ($-125^\circ, \ -18^\circ$)} & \makecell{700 \\ 1085} & \makecell{$-54^\circ$ \\ $41^\circ$} & \makecell{($53^\circ, \ 26^\circ$) \\ ($161^\circ, \ 51^\circ$)} \\ \hline
	
	2014-06-10 16:00 & \makecell{pre 11:48 \\ pri 13:30} & \makecell{87 \\ Halo} & \makecell{925 \\ 1469} & \makecell{$111^\circ$ \\ $360^\circ$} & \makecell{($-106^\circ, \ -16^\circ$) \\ ($-105^\circ, \ -20^\circ$)} & \makecell{888 \\ 1110} & \makecell{$-23^\circ$ \\ $-76^\circ$} & \makecell{($101^\circ, \ 38^\circ$) \\ ($177^\circ, \ 57^\circ$)} \\ \hline
	
\end{longtable}
\begin{tablenotes}
	\item[1] [1] The table contains the SEP onset time and the related information of the preCME and priCME.	
	\item[2] [2] 1st App. means the CME first appearance time in LASCO/C2 FOV, $v$ is the CME speed in the unit of km/s, $cpa$ and $w$ are the central position angle and angular width of a CME in the CDAW/CME catalog; the GCS model results: $(\phi, \theta)$ are the longitude and latitude of the propagation direction, $\gamma$ is the tilt angle, and $(w_f, \ w_e)$ refer to the GCS face-on and edge-on angular widths.
	\item[3] [3] The superscript of ``$-1$'' indicates that one day prior to the SEP onset day. 
\end{tablenotes}
\normalsize

\begin{figure}[!hbt]
	\centering
	\includegraphics[width=0.9\textwidth]{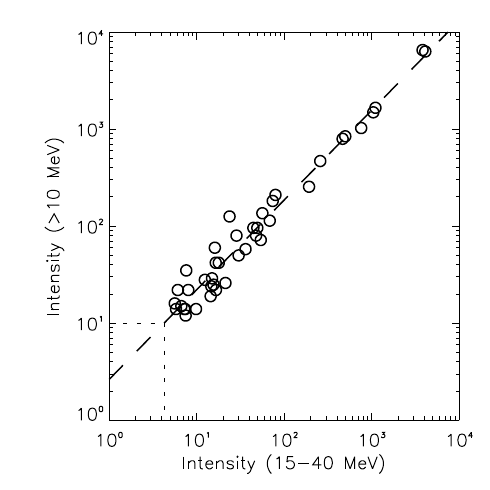}
	\caption{\small Plots of the SEP peak intensities observed by GOES in the energy ranges of 15--40 MeV vs. $>10$ MeV. The dashed line indicates the linear-fit to the data points, and the dotted lines mark the threshold for large SEP events. } 
	\label{determine_thres}
\end{figure}
Figure \ref{determine_thres} shows the SEP peak intensities observed by GOES in the energy ranges of 15--40 MeV vs. those $>10$ MeV for the large SEP events in SC24 listed in the CDAW/SEP catalog. This process is needed to identify the threshold for events best measured by STEREO due to our usage of the 15--40 MeV energy range rather than $>10$ MeV for events measured by STEREO. It is found that the intensities in both energy ranges are correlated with correlation coefficient (CC) equaling 0.982. The dashed line in the figure indicates the linear-fit to the data points, and then the threshold of SEP-rich events in 15--40 MeV corresponding to NOAA definition of $>10$ pfu for protons $>10$ MeV can be determined as the related peak intensity reaches 4.3 pfu (indicated by the vertical dotted line); the peak intensity for SEP-poor events in 15--40 MeV is $<4.3$ pfu.

It is well known that the magnetic connectivity between the observation point and the SEP source region has significant influence on the SEP observation \citep[e.g.,][]{2013ApJ...767...41L,2014EP&S...66..104G,2014SoPh..289.3059R,2016ApJ...819...72L,2019JGRA..124.6384X}. In this work, we use the fitting equation provided by \citet{2019JGRA..124.6384X} to estimate the peak intensity at a point with the best magnetic connectivity ($I_0$). The equation is re-written as $I(CA)=I_0 \exp (- CA^{2} / 2 \sigma^{2})$, where $CA$ is the connection angle with respect to a spacecraft, i.e., the longitudinal separation angle between the SEP source region and the magnetic footpoint of the observing spacecraft. Positive $CA$ denotes a SEP source region located at the western side of the spacecraft magnetic footpoint; negative $CA$ represents a source to the east. The longitude of the magnetic footpoint is calculated by assuming a nominal Parker spiral. In the equation of \citet{2019JGRA..124.6384X}, $\sigma=8.4+0.21CA$ if $CA \ge 0$, or (2) $\sigma=7.9-0.28CA$ if $CA < 0$. It reflects an east-west asymmetry of the longitudinal distribution with the $CA$-dependent distribution width ($\sigma$), and the intensity is highest at zero $CA$. The effectiveness of the east-west asymmetric SEP distribution was discussed by \citet{2019JGRA..124.6384X}. It shall be noted that in some cases, the flank of the shocks may be the locations where the acceleration is most efficient \citep[e.g.,][]{2019ApJ...876...80K}, and this strongest portion may not longitudinally coincide with source region at $CA=0^\circ$. For every event, we input $CA$ and $I_p$ at the spacecraft with the highest $I_p$ to estimate $I_0$, and then the related longitudinal distribution of $I(CA)$ can be obtained accordingly. Noted that \citet{2019JGRA..124.6384X} derived the fitting equation based on 20--30 MeV proton energy range, and here we directly use the 15--40 MeV $I_p$ as a nearly linear relationship can be expected for $I_p$ in these two ranges (as seen in Figure \ref{determine_thres}) for 15--40 MeV protons.

\begin{figure}[!hbt]
	\includegraphics[width=\textwidth]{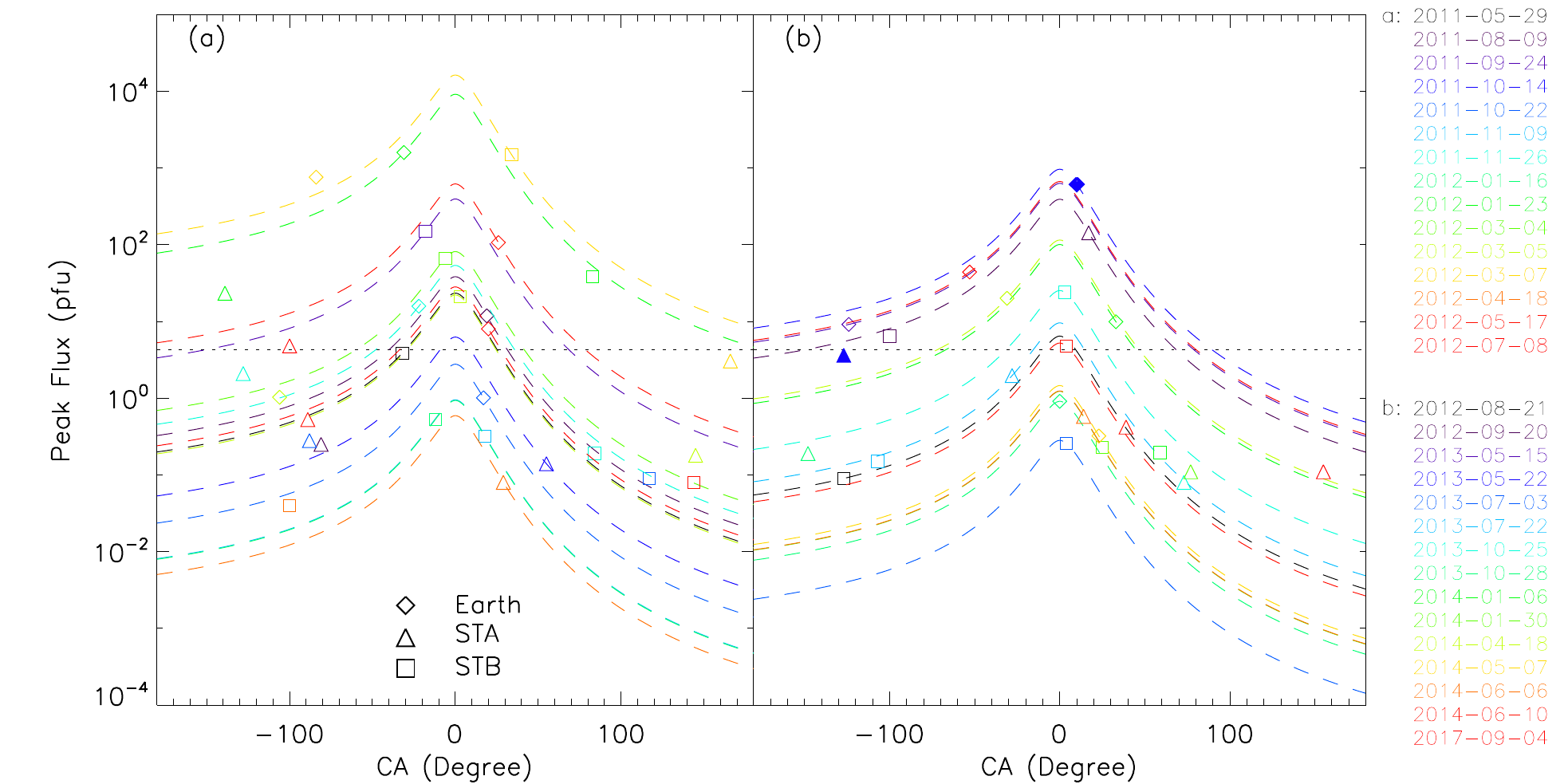}
	\caption{\small Estimated longitudinal distribution of $I_p$ as the function of connection angle ($CA$). The observed values are plotted by the triangle (STA), diamond (GOES) and square (STB), respectively. The horizontal line indicates the 4.3 pfu threshold for large SEP events. $I_0$ as listed in Table \ref{sep_ip} and \ref{large_sep} correspond to the peak of each distribution.} 
	\label{int_lon}
\end{figure}

Figure \ref{int_lon} shows the estimated $I_0$ and the longitudinal distribution of $I(CA)$ (dashed curve), associated with the observed $I_p$ for the SEP events in Table \ref{sep_ip} and \ref{large_sep}.
It is found that the observed values from other spacecraft can not always match well with the related curves (e.g., see one case on 2013 May 22 with filled symbols in Figure \ref{int_lon}(b)), but these inconsistent data points are almost in poorly magnetic-connected region. Thus, the $I_0$ is still effective for a reference, and the related values are listed in the last column in the Table \ref{sep_ip} and \ref{large_sep}. Besides, using nominal Parker spiral field line can lead to a $\sim 20$--$30^\circ$ uncertainty in the estimation of the magnetic footpoint, which may influence the results in Figure \ref{int_lon}. Based on this figure, we divide the 19 L1-point SEP-poor events into three groups: (1) Group-I, with at lease one of the observed $I_p$ $>4.3$ pfu (six events), (2) Group-II, with only $I_0>4.3$ pfu but not observations at $I_p>4.3$ pfu (four events), and (3) Group-III, with $I_0<4.3$ pfu (nine events). The 11 events in Table \ref{large_sep} are catagorized in Group-I. One can refer to the Appendix for a further discussion about using the equation by \citet{2019JGRA..124.6384X}.  

\subsection{Statistical Analysis}\label{sec_sta}
In this section, we study the statistical relationships between the SEP $I_p$ on the one hand, and the source location, the CME speed and propagation direction in latitude, the flare size, the seed populations, and the CME-CME interaction on the other hand. The events in Table \ref{sep_ip} and \ref{large_sep}, i.e., SEP-poor and SEP-rich events as measured from L1 point, are analyzed together. 

\subsubsection{SEP Source Location Distribution}
\begin{figure}[!hbt]
	\centering
	\includegraphics[width=0.95\textwidth]{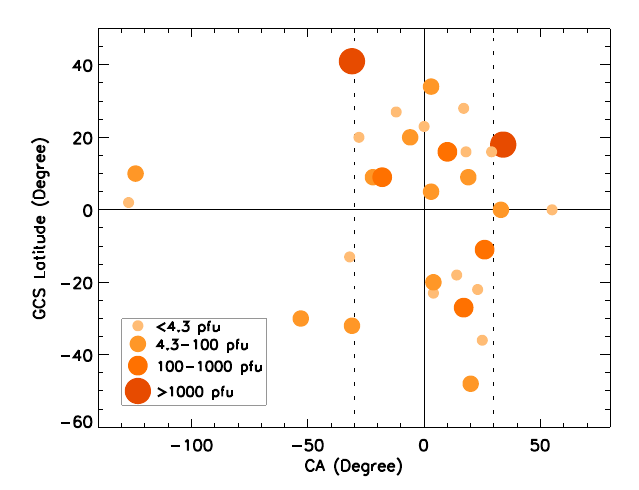}
	\caption{\small Distribution of the source region of the SEP events in $CA$-latitude space. The size and darkness of the circles correspond to the related $I_p$ observed by the spacecraft with the best magnetic connectivity.} 
	\label{caip}
\end{figure}

Figure \ref{caip} shows the distribution of the source region of the SEP events in $CA$-latitude space, in which the size and darkness of the circles correspond to the $I_p$ level observed by the spacecraft with the best magnetic connectivity.
It is found that the samples are distributed in random latitudes (though still not too far away from $0^\circ$), and most of the data points lie in or close to the magnetic well-connected region ($-30^\circ \le CA \le 30^\circ$). There are two special data points very far away from the well-connected region, which correspond to the events on 2012 August 21 and 2013 May 15, respectively. The observations of these events at the spacecraft with the best magnetic connectivity are not used because they are saturated by the preceding events. This figure illustrates the advantage of making this study in SC24 when the twin STEREO spacecraft were available (2010--2014) and were $\sim80$--$180^\circ$ from Earth. As such, there was a great coverage of the $2\pi$ heliosphere, and this is why most events have one spacecraft within $\pm 30^\circ$ from the central meridian longitudes. However, even with observations with $|CA| \le 30^\circ$, there still exist 11 twin-CME events in which the observed SEP peak intensity is lower than the threshold for large SEP events.

\subsubsection{CME Speed vs. SEP Peak Intensity}
\begin{figure}[!hbt]
	\centering
	\includegraphics[width=0.98\textwidth]{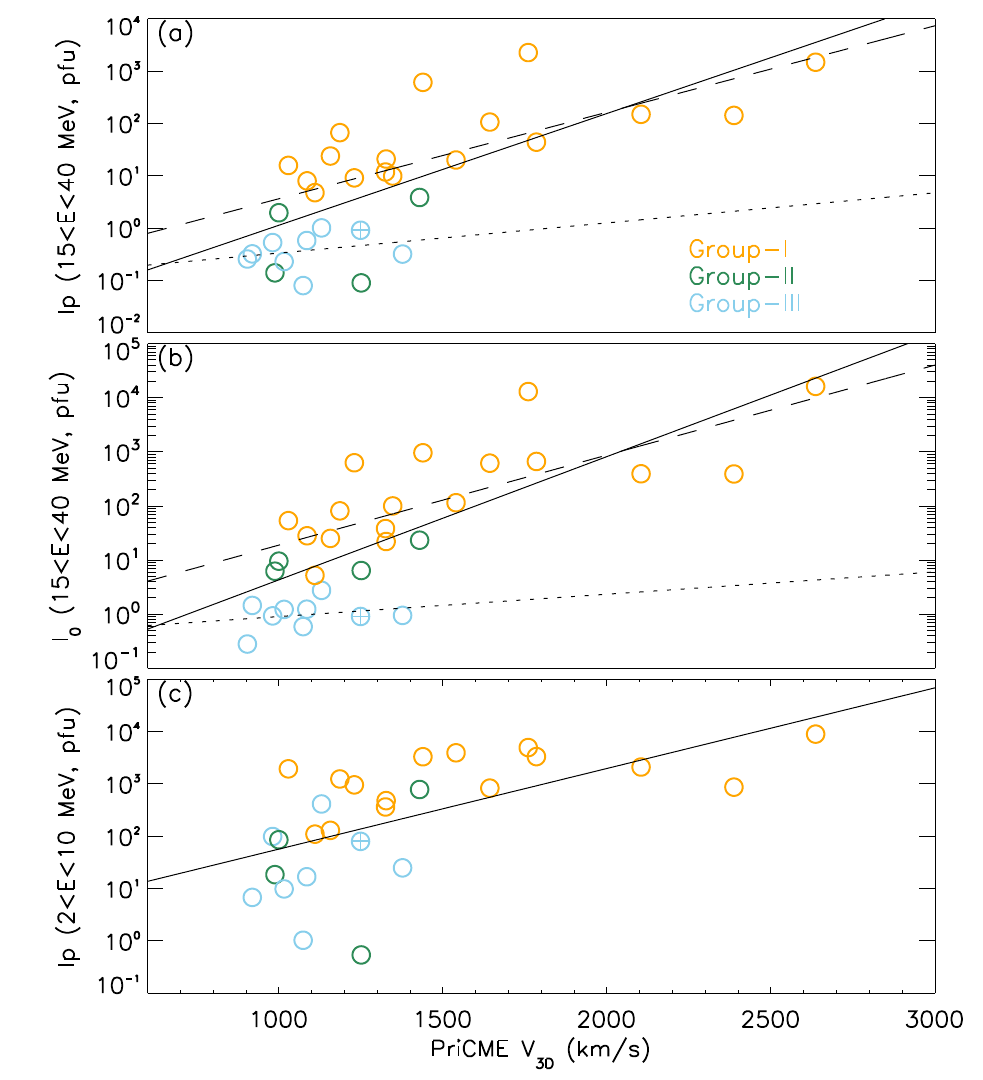}
	\caption{\small Correlations between the SEP peak intensity and the priCME 3D speed. The data points in the three groups are shown by different colors. The solid, dashed and dotted lines indicate the linear-fit to the clusters of all the events, SEP-rich events and SEP-poor events, respectively.} 
	\label{priCME_SEP}
\end{figure}

Figure \ref{priCME_SEP} provides the correlations between the SEP peak intensity and the priCME 3D speed ($V_{pri}$). 
Panel (a) shows the data points of 15--40 MeV proton $I_p$ vs. $V_{pri}$ obtained from the GCS reconstruction, and there is strong correlation with ${\rm{CC}}\ (\log I_p \ vs. \ V_{pri})=0.726$ when considering all the data points. However, if the correlations are analyzed in two types of clusters, i.e., one with the events in Group-I and II (SEP-rich events), and the other with the events in Group-III (SEP-poor events), then it is found that (1) CC for the SEP-rich events decreases to 0.675, and (2) CC for the SEP-poor events is lower as 0.261. Panel (b) shows the correlations of $I_0$ vs. $V_{pri}$, and the CC for the SEP-rich events is found to be slightly higher than CC of $I_p$ vs. $V_{pri}$ for the SEP-rich events (0.754 vs. 0.675). It may indicate the role of correcting the longitudinal distribution in increasing the correlations even we have used the observations from the best-connected spacecraft. However, CC for the SEP-poor events is still low as 0.233. Panel (c) gives the correlation of $I_p$ in 2--10 MeV vs. $V_{pri}$ (note that there are three events without the observations in this energy range), in which it is found that ${\rm{CC}}=0.594$, 0.521 and 0.259 for all the events, SEP-rich and SEP-poor events, respectively. Comparisons of CCs between all the events and the SEP-rich events highlight the importance of including SEP-poor events to really understand how SEP peak intensity relates to CME speed.

\subsubsection{Influence of CME Latitudinal Propagation Direction on SEP Peak Intensity}\label{lateffect}
The effect of the CME propagation direction in latitude is considered next. It is easy to imagine that if the CME (or the source region) is out of the ecliptic plane, then the SEP enhancement as measured by spacecraft in the ecliptic becomes weaker (one can also see \citet{2010AIPC.1216..613D} for the influence of the source regions in latitude on SEP detection). In this work, the latitudinal effect is considered through two types of corrections. The first is that $V_{pri}$ is projected to the ecliptic plane by multiplying a factor of $\cos (\triangle \theta)$, where $\triangle \theta$ is the angular separation between the latitude of the CME propagation direction and the Sun-Earth line (the separation between the solar equatorial plane and the ecliptic plane varies from $\sim 7^\circ$ to $\sim -7^\circ$ throughout an Earth year). This projection is based on the assumption that the SEPs accelerated in the ecliptic plane are related to the CME in the same plane. The second is that the peak intensities are corrected by multiply a factor of $\exp({-|\triangle \theta|/\theta_c})$, or $V_{pri}$ is divided by $\exp({-|\triangle \theta|/\theta_c})$, according to the equation provided by \citet{1997JGR...10222335K}, where $\theta_c$ is a constant. This correction indicates that the latitudinal effect acts somewhat like that the longitudinal effect. We set $\theta_c$ as $20^\circ$ to make the correlations have the highest CCs in general. The related results are given in Table \ref{cme_sep_cc} (the plots are similar to Figure \ref{priCME_SEP} and not shown here), in which it is found that the corrections do not increase the related CCs, especially for the first correction. However, the second increases CC for the SEP-poor events significantly, which indicate that the latitudinal effect is important in that group. The decreases in the SEP-rich group may be due to the fact that (1) the real latitudinal effect is different from the above expectations which would be the case in particular if the shock shape is more complex and can not be represented by a simple cosine function, or (2) the background Alfv\'{e}n speed, which is one of the dominant factors in determining the shock strength, may differ significantly in latitude.

\begin{table*}[!htb]
	\footnotesize
	\centering
	\caption{Correlation coefficients of SEP peak intensity vs. CME speed in different types. }
	\begin{tabular}{|c|c|c|c|c|c|c|c|c|c|}
		\hline
		&  \multicolumn{6}{c|}{15--40 MeV} & \multicolumn{3}{c|}{2--10 MeV} \\ \cline{2-10}
		\diagbox{CME}{SEP} & \multicolumn{3}{c|}{$\log I_p$} & \multicolumn{3}{c}{$\log I_0$} & \multicolumn{3}{|c|}{$\log I_p$} \\ \cline{2-10}
		& All & Rich & Poor & All & Rich & Poor & All & Rich & Poor \\ \hline
		$V_{pri}$ & 0.726 & 0.675 & 0.261 & 0.756 & 0.754 & 0.233 & 0.594 & 0.521 & 0.259 \\ \hline
		$V_{pri}\cos(\triangle \theta)$ & 0.626 & 0.545 & 0.234 & 0.674 & 0.643 & 0.216 & 0.521 & 0.421 & 0.172 \\ \hline
		$V_{pri}/ \exp({-|\triangle \theta|/\theta_c})$ & 0.711 & 0.650 & 0.473 & 0.726 & 0.708 & 0.476 & 0.559 & 0.452 & 0.373 \\ \hline
		$E_k$ & 0.625 & 0.566 & 0.266 & 0.615 & 0.569 & 0.540 & 0.470 & 0.388 & 0.221 \\ \hline
	\end{tabular}	
	\label{cme_sep_cc}
	\begin{tablenotes}
		\item[1] [1] ``All'', ``Rich'' and ``Poor'' indicates there groups with all the events, SEP-rich events and SEP-poor events, respectively.
	\end{tablenotes}
\end{table*}
\normalsize

Overall, we find that: (1) there exist correlations between $V_{pri}$ and SEP $I_p$, both in higher and lower energy ranges (a linear correlation also exist between these two ranges with ${\rm{CC}}=0.883$); (2) the angular separation between the SEP source and spacecraft in longitude shall be considered for obtaining a better correlation; and (3) the CMEs with $V_{pri}\gtrsim 1500$ km/s drive SEP-rich events but those $V_{pri}\lesssim 1500$ km/s can drive either SEP-rich or SEP-poor events. The linear fits to the data points for all the events (solid line), the SEP-rich events in Group-I and II (dashed line), and the SEP-poor events in Group-III (dotted line) are shown in Figure \ref{priCME_SEP}. It is found that the slopes of the dashed lines are similar to those of the solid lines, but different from those of the dotted lines, indicating that, statistically, other factors, e.g., the latitude of the CME propagation direction, may be as important as the CME speed for the SEP-poor events. Furthermore, following \citet{2019JGRA..124.6384X}, we also analyze the relationship between the CME kinetic energy ($E_k$) and the SEP peak intensity, where the $E_k \propto v^2 \kappa^2 \omega_{f}$ based on the GCS model results and the uniform-density assumption \citep[see][]{2019JGRA..124.6384X}. The corresponding CCs are also given in Table \ref{cme_sep_cc}, and it is found that the CCs become lower, except one (${\rm{CC}}=0.540$ for $E_k$ vs. $\log I_0$ in 15--40 MeV) in the SEP-poor group.

The different properties about the relationship in SEP-rich and SEP-poor events, or the general CME-SEP relationship, may also be influenced by various factors. In the following part, we analyze the roles of the flare size, suprathermal ion backgrounds, and the effects of the CME-CME interaction in the SEP enhancements. 

\subsubsection{Flare Size and Suprathermal Backgrounds vs. SEP Peak Intensity}
\begin{figure}[!hbt]
	\includegraphics[width=\textwidth]{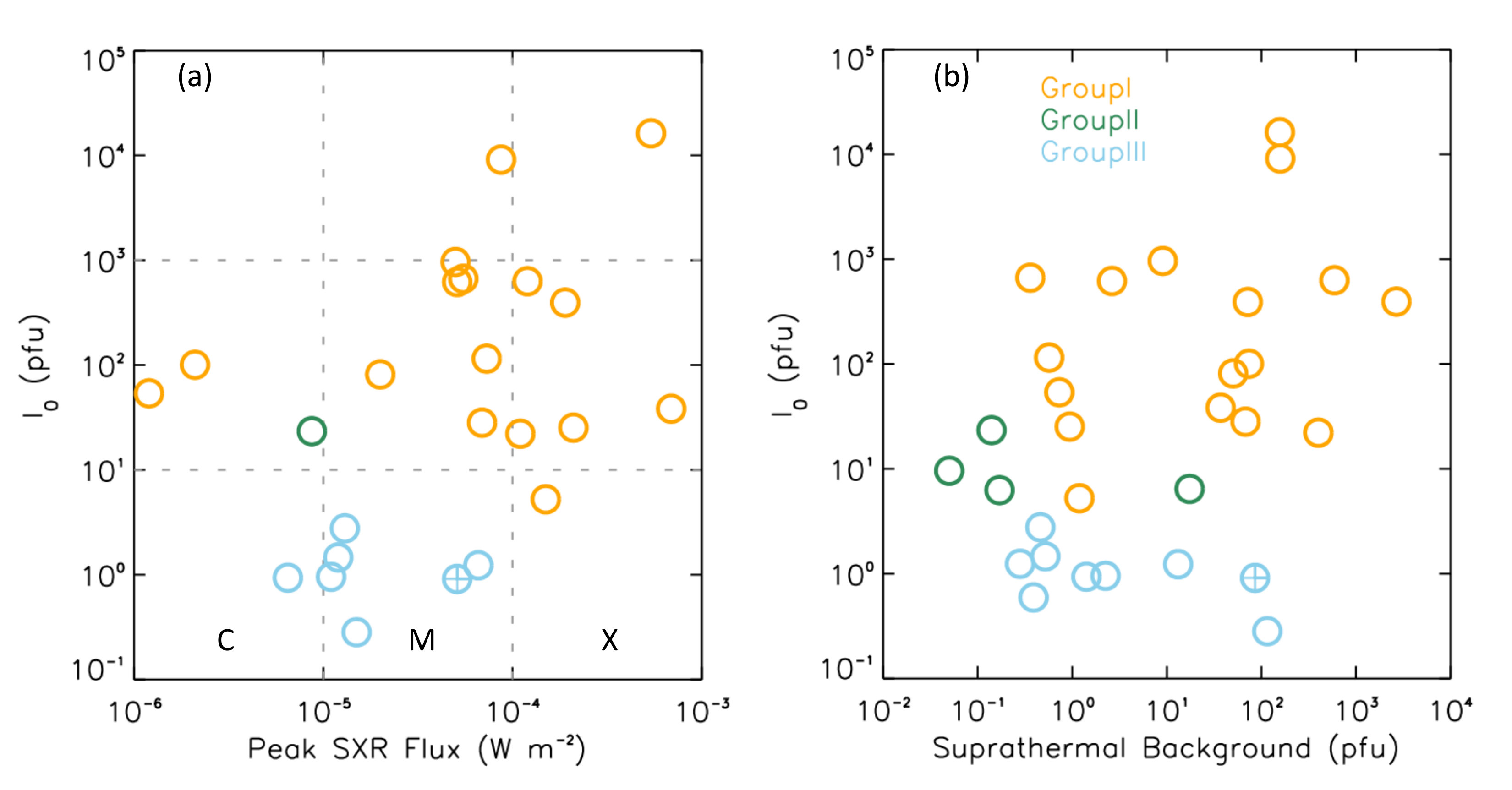}
	\caption{\small (a) Plot of the soft X-ray peak flux vs. SEP peak intensity for the events with GOES observations. (b) Plot of the 1--2 MeV suprathermal ion background in a 4-hour duration before the SEP onset vs. SEP peak intensity.} 
	\label{flare_back_ip}
\end{figure}

As for the role of flare in contributing to SEP acceleration, Figure \ref{flare_back_ip}(a) shows the plot of the SXR peak flux ($I_{SXR}$) vs. $I_0$ for the events with GOES SXR observations (24 events). There exists a weak correlation for all the data points with ${\rm{CC}} \ (\log I_{SXR} \ vs. \ \log I_0)=0.384$, and the events with higher $I_0$ tend to be associated with stronger flares. Our CC is close to $\rm{CC}=0.40$ reported in \citet{2020ApJ...900...75K} as they studied 31 events with source regions in the western hemisphere. However, due to the close relationship between flares and CMEs, this flare-SEP correlation may be intrinsically related to the CME-SEP correlation. Refer to Section \ref{sec_multi_role} for a further discussion.

As for the suprathermal proton background, following \citet{2019ApJ...872...89K}, we calculate the averaged values of the 1--2 MeV proton intensity ($I_{b}$) in four periods, 24-hour before, 4-hour before, 4-hour after and 8-hour after the SEP onset, respectively. Only the intensities at the spacecraft with the best connectivity are used. This assumes that the suprathermal protons observed at the spacecraft are representative a population that extends along the magnetic field lines back to the corona, and an 1--2 MeV proton takes around 170 minutes to propagate from the lower corona to 1 AU along the nominal Parker spiral with the length of 1.15 AU. Figure \ref{flare_back_ip}(b) shows the plot of $I_b$ four hours before the SEP onset vs. $I_0$. Table \ref{supra_sep_cc} lists the corresponding CCs in different conditions, and the CCs for the $>10$ MeV SEP events within $0^\circ$--W$40^\circ$ and W$41^\circ$--W$83^\circ$ from \citep{2019ApJ...872...89K} are also listed in parentheses for comparison. It is found that the coefficients in the SEP-rich group for the four periods are around 0.5, indicating a correlation between $I_b$ and $I_0$. As for the SEP-poor events in the last column, the analyses by the student's $t$-test show that there is no $I_b$-$I_0$ correlation. The comparisons of the CCs between ours and from \citet{2019ApJ...872...89K} show that our CCs are consistent with Kahler's for suprathermal backgrounds averaged in the four periods within $0^\circ$--W$40^\circ$ and in the two after-onset periods within W$41^\circ$--W$83^\circ$. However, CCs from \citet{2019ApJ...872...89K} for the two before-onset periods within W$41^\circ$--W$83^\circ$ are significantly lower. In general, W$41^\circ$--W$83^\circ$ source regions are supposed to be magnetic well-connected with the spacecraft near the Earth, while \citet{2019ApJ...872...89K} used the in situ data from L1 point. The inconsistency between our samples which are almost within $-30^\circ \le CA \le 30^\circ$ and Kahler's within W$41^\circ$--W$83^\circ$ may be due to the fact that the observations of the suprathermal populations before the SEP events are limited in a narrow longitudinal range.

\begin{table*}[!htb]
	\footnotesize
	\centering
	\caption{Correlation coefficients of suprathermal background vs. SEP peak intensity (in log-scale). }
	\begin{tabular}{|c|c|c|c|}
		\hline
		\diagbox{Suprathermal}{SEP} & All & \makecell{Rich \\ (Kahler: $0^\circ$--W$40^\circ$, W$41^\circ$--W$83^\circ$)} & Poor \\ \hline
		24 hr before & 0.359 & 0.446 (0.480, 0.120) & -0.561 \\ \hline
		4 hr before & 0.388 & 0.458 (0.500, 0.050) & -0.561 \\ \hline
		4 hr after & 0.573 & 0.516 (0.530, 0.360) & -0.534 \\ \hline
		8 hr after & 0.642 & 0.573 (0.630, 0.560) & -0.454 \\ \hline
	\end{tabular}	
	\label{supra_sep_cc}
	\begin{tablenotes}
		\item[1] [1] ``All'', ``Rich'' and ``Poor'' indicates there groups with all the events, SEP-rich events and SEP-poor events, respectively. CCs in \citet{2019ApJ...872...89K} are listed in parentheses.
	\end{tablenotes}
\end{table*}
\normalsize

\subsubsection{The Influence of CME-CME Interaction on SEP Peak Intensity}
\begin{figure}[!hbt]
	\includegraphics[width=\textwidth]{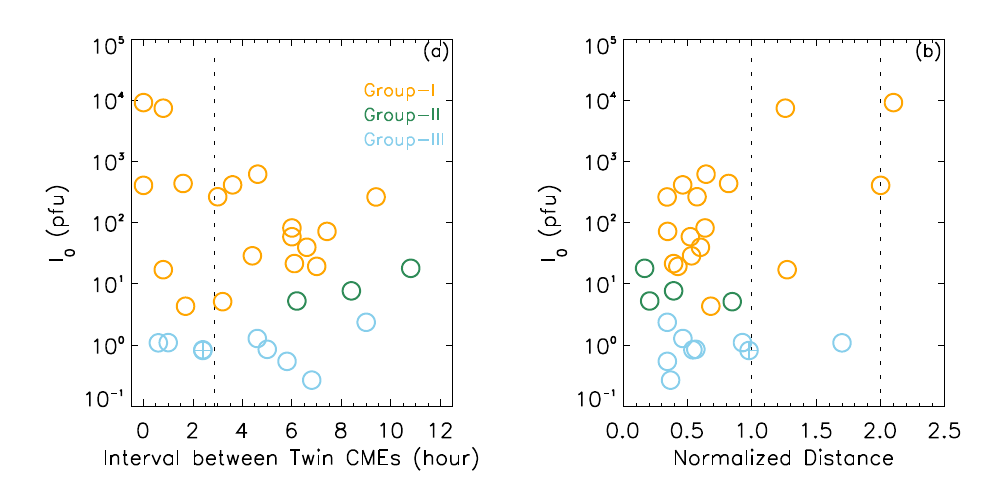}
	\caption{\small (a) Plot of the time interval between the twin CMEs vs. $I_0$, and the vertical line indicates the 1--2 MeV proton transportation time (170 minutes). (b) Plot of $d_n$ vs. $I_0$. The two vertical dotted lines indicate the magnetic ejecta region of the preCME.} 
	\label{inte_ip}
\end{figure}

As for the effect of the CME-CME interaction, we first show the time interval of the eruption between twin CMEs vs. $I_0$ in Figure \ref{inte_ip}(a) (the intervals of two special events on 2012 March 7 and 2012 May 17 are set to be zero hour). 
The eruption time is simply estimated as the time of first appearance time in projected LASCO/C2 FOV and the CME 2D speed. It is found that for both the SEP-rich and SEP-poor events associated with successive CMEs, $I_0$ is independent on the time interval between the eruptions. In the twin-CME scenario, preCMEs may play a role in providing enhanced seed populations and turbulence level. However, it is hard to estimate the turbulence level in the corona, and the observations of the seed populations are limited by the propagation effect. This is because 1--2 MeV protons needs $\sim 170$ minutes to propagate from the Sun to a spacecraft at 1 AU along the nominal Parker spiral which requires that the interval shall be at least greater than 170 minutes. The vertical dotted line in Figure\ref{inte_ip}(a) marks this threshold, and nearly one third of our samples do not meet this requirement. Although we have shown a weak correlation between the suprathermal protons in a 4-hour period before the SEP onset and $I_0$ in Table \ref{supra_sep_cc}, we follow \citet{2020ApJ...901...45Z} to use the normalized distance ($d_n$) at the particle release time to indicate the effect of the CME-CME interaction. $d_n$ is expressed as $d_n=1+\frac{h_{LE-priCME}-h_{TE-preCME}}{2R_{preCME}} \ge 1$ if the priCME impacts the preCME, or $d_n=\frac{h_{LE-priCME}}{h_{TE-preCME}}<1$ on the contrary. $h_{LE-priCME}$ is the height of the priCME leading edge, $h_{TE-preCME}$ is the height of the preCME trailing edge, and $R_{preCME}$ is the preCME radius, which are obtained from the GCS model. $d_n=1$ refers to the condition that the leading edge of priCME rightly catches up with the trailing edge of preCME, while $d_n=2$ is for the condition that the leading edges of the two CMEs have fully merged. In this paper, $d_n$ is roughly estimated by the assumption that the particles are released when the priCME reaches the height of 4 Rs \citep[see][]{2012SSRv..171..141L}. Figure \ref{inte_ip}(b) gives the plot of $d_n$ vs. $I_0$, and the ${\rm{CCs}} \ (d_n \ vs. \ \log I_0)$ are found to be 0.357, 0.569 and 0.178 for all the events, SEP-rich and SEP-poor events, respectively. It indicates the considerable effect of the CME-CME interaction for at least the SEP-rich events, and it also shows that the meaningful CME-CME interaction is not necessary for the particle release in the twin-CME scenario in agreement with in \citet{2020ApJ...901...45Z}.

\section{Discussion}\label{sec_dis}
\subsection{Multiple Roles in the SEP Enhancement}\label{sec_multi_role}
So far, it has been found that numerous factors influence the peak intensity of SEP events (simultaneously). Our  analyses presented above rely on one-on-one correlations. In this section, we investigate the relationship between multiple independent factors together and one dependent variable of $I_0$. These independent factors include the priCME 3D speed ($V_{pri}$), the latitude of the priCME propagation direction with respect to the ecliptic plane ($\triangle \theta$), the flare size ($I_{SXR}$), the suprathermal ion backgrounds in a 4-hour period before the SEP onset ($I_b$), the preCME speed ($V_{pre}$), and the normalized distance ($d_n$) in CME-CME interaction. Based on the statistics in Section \ref{sec_sta}, we presume $V_{pri}$ as the major independent variable because the related CCs are at the highest level among these parameters. Before analyzing the correlations, the relationship between some of the independent variables is studied at first because they belong to the chain of solar activities/eruptions \citep[see a recent review paper][]{2020arXiv201206116Z}, and the correlation effect between these variables themselves may be involved.

Many papers have revealed the close relationship between the CME and flare \citep[e.g.,][]{2009IAUS..257..233Y,2010ApJ...712..752C,2020ApJ...897L..36G,2020arXiv201206116Z}, and the related correlation coefficients are also given by, e.g., 0.50 in \citet{2009IAUS..257..233Y}, or 0.47 in \citet{2010ApJ...712..752C}. Our samples reveal that ${\rm{CC}} \ (V_{pri} \ vs. \log \ I_{SXR})=0.470$, indicating a moderate correlation between the CME speed and the flare size. The flare-SEP correlation in Figure \ref{flare_back_ip}(a) may intrinsically caused by the stronger CME-SEP correlation. As hypothesized by the twin-CME scenario \citep{2012SSRv..171..141L}, the flare materials (accompanying both the preCME and the priCME) could be released into the upstream of the second shock driven by the priCME via interchange reconnection as seed populations. The enriched $\rm{^3 He}$ intensity during the large SEP events observed by in situ spacecraft also indicates that the residual or remnant flare-accelerated $\rm{^3 He}$-rich suprathermal material can contribute to the seed populations accelerated by CME-driven shock \citep[e.g.,][]{1999ApJ...525L.133M,2012SSRv..171...97M}. Therefore, the role of the flare size is still analyzed here. It shall be noted that the poor $I_{SXR}$-SEP peak relationship found by \citet{2014JGRA..119.9456P} and \citet{2015SoPh..290..819T} may be due to the fact that the flare size is not related to the interchange reconnection process which plays a major role in the release of seed populations \citep[e.g.,][]{2013ApJ...771...82M,2019ApJ...884..143M}. 

The role of suprathermal backgrounds (or seed populations) was previously discussed in \citet{2013ApJ...763...30D} and \citet{2014ApJ...784...47K}. \citet{2013ApJ...763...30D} found that the 1-MeV proton intensities in a 24-hour duration before the SEP do not correlate with the SEP peak intensities for the twin-CME events. Their interpretation is that the preCMEs can act on the newly produced seed particles and thus the SEP peak would be independent of the backgrounds. However, \citet{2014ApJ...784...47K} argued that the level of seed populations is only solar activity dependent. Here we consider role of the suprathermal backgrounds no matter whether it act through the twin-CME or higher level of solar activity.

In Table \ref{supra_sep_cc}, CC between $I_0$ and $I_b$ increases when using the background intensities after the SEP onset. It is found that CCs of $V_{pri}$ vs. $\log I_b$ are 0.384, 0.417, 0.492, and 0.519 from the first to the fourth period, respectively. This may indicate that the higher CCs for the after periods are caused by considering the 2-MeV particles accelerated by the priCMEs (recalling the 170-minute transport time). As for $V_{pre}$ vs. $\log I_b$, CCs are found to be low as $\sim 0.3$ for the second to the fourth period, indicating that the level of the seed populations may not be related to the preCMEs (or at least to their speed) for our samples.

Since we need to consider all the independent variables simultaneously, we then use the projection correlation method \citep{zhu2017projection,liu2020model} to analyze the correlations between the combination of multiple independent variables and one dependent variable. This method has four properties: (1) the correlation coefficient equals zero only if the independent variable(s) and the dependent variable(s) are independent, (2) there is no limitation in the variable dimensions, (3) the coefficient is independent on the group of orthogonal transformations, and (4) the estimation does not require specific tuning parameters and moment conditions on the variables. In the following part, the coefficient of the projection correlation is named as PC, and CC still represents the linear Pearson correlation coefficient.
\begin{figure}[!hbt]
	\includegraphics[width=\textwidth]{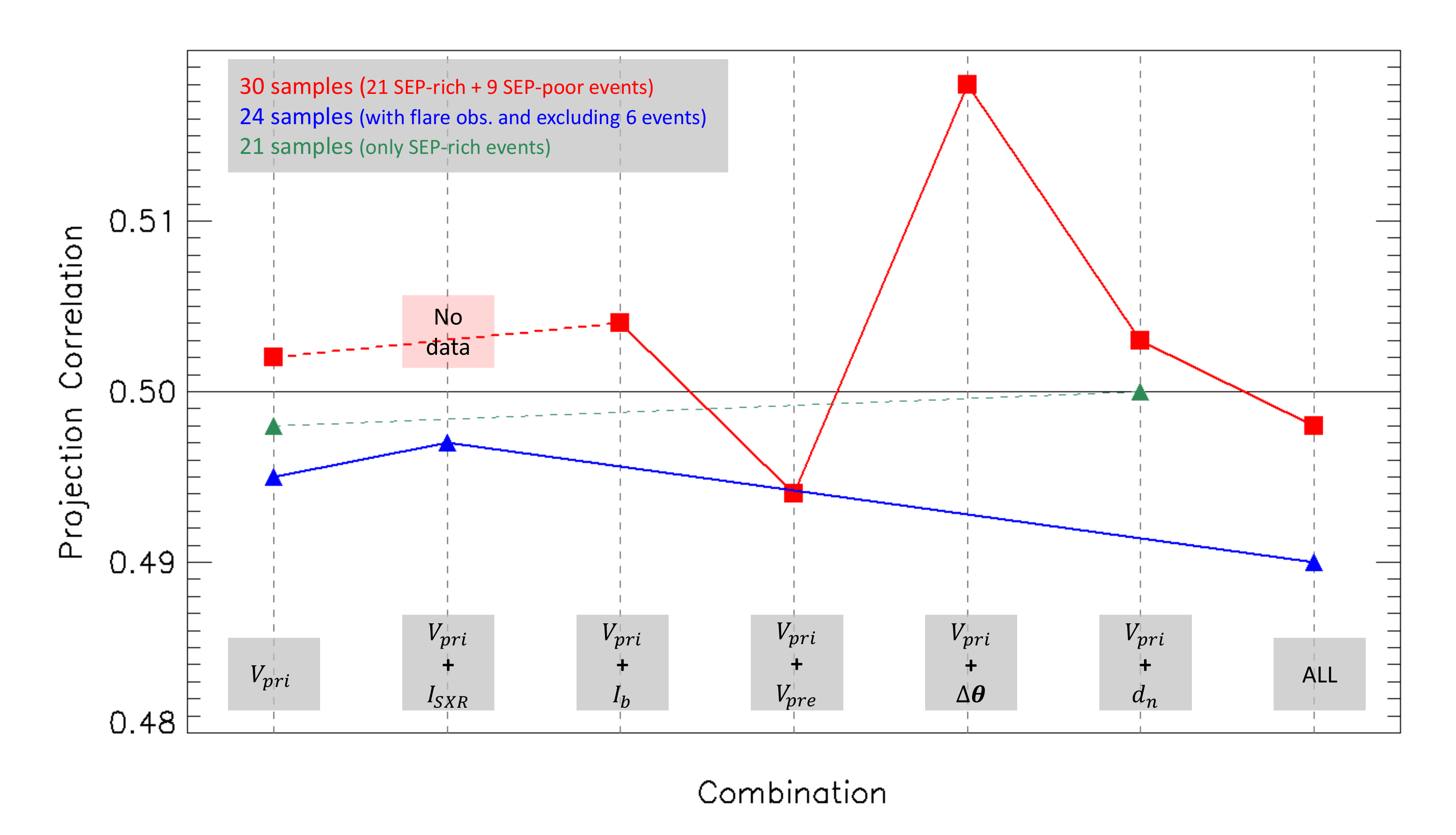}
	\caption{\small Projection correlation coefficients between SEP $I_0$ and different combinations of independent variables. Three groups of the samples are considered: (1) all the 30 events not associated with flare information (red), (2) 24 events associated with flare information by excluding six events which do not have GOES SXR observations (blue), and (3) 21 SEP-rich events not associated with flare information (green).} 
	\label{projection_c}
\end{figure}

We analyzed the samples in three groups, (1) all the events not associated with flare information (30 events), (2) 24 events associated with flare information (six events do not have GOES SXR observations), and (3) the SEP-rich events not associated with flare information (21 events). The PCs in different combination conditions are shown in Figure \ref{projection_c}, in which the dashed line indicates that there is no data for its crossed combination, and PC above 0.5 indicates a quite good correlation in this method. These combinations contain (1) $V_{pri}$ vs. $I_0$, (2) ($V_{pri},\ I_{SXR}$) vs. $I_0$, (3) ($V_{pri},\ I_b$) vs. $I_0$, (4) ($V_{pri},\ V_{pre}$) vs. $I_0$, (5) ($V_{pri},\ \triangle \theta$) vs. $I_0$, (6) ($V_{pri},\ d_n$) vs. $I_0$, (7) ($V_{pri},\ I_b,\ V_{pre},\ \triangle \theta,\ d_n$) vs. $I_0$ for the red and green data points, and (8) ($V_{pri},\ I_{SXR},\ I_b,\ V_{pre},\ \triangle \theta,\ d_n$) vs. $I_0$ for the blue data points, respectively. Different correlation methods may lead to different coefficient values. We take $V_{pri}$ as the dominant role, and the PCs in the first column act as a reference for the following combinations. It is found that for the three groups of samples, the combinations do not lead to relatively and visibly higher PCs except the one with ($V_{pri},\ \triangle \theta$). This increase indicates that the consideration of the latitude of the CME propagation direction (or the source region similarly) can lead to a better understanding about the CME-SEP relationship, which further supports the result in Section \ref{lateffect}. Although we do not know the exact form of the latitudinal effect compared to the longitudinal effect (considering the different coronal magnetic field topology in latitude and longitude), we find a rough assumption of the exponential distribution results in the increase of CC for the SEP-poor events as shown in Table \ref{cme_sep_cc}. Further analyses will be made possible with the observations from the Parker Solar Probe and the Solar Orbiter. Figure \ref{projection_c} also supports the dominant role of the priCMEs in controlling the SEP enhancement, while the roles of the flare size, suprathermal backgrounds, the preCMEs, or the CME-CME interaction (though a quite strong correlation of $d_n$ vs. $I_0$ is found for the SEP-rich events in Figure \ref{inte_ip}) incorporated do not have additional contribution to SEP peak intensities. This may be due to the fact that (1) these factors are closely coupled with priCMEs (such as the flare-CME coupling), or (2) their roles are more complicated than what we are able to investigate in this study as we are limited to a relatively moderate number of events in SC24 since we rely on the STEREO observations. 

\subsection{Specific Event on 2013 October 28}\label{sec_20131028}
\begin{figure}[!hbt]
	\includegraphics[width=\textwidth]{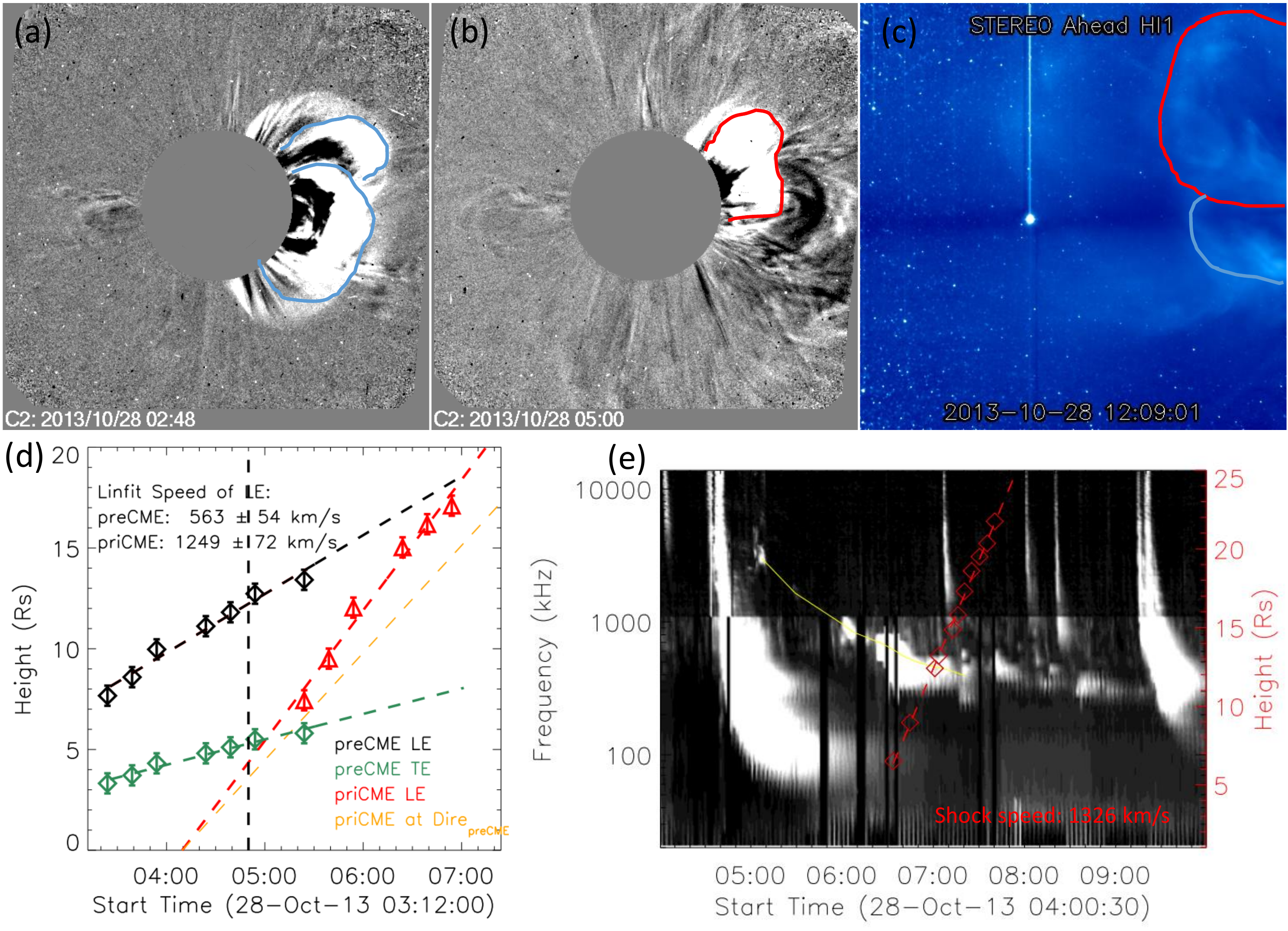}
	\caption{\small (a)--(c) The preCMEs (blue curve) and priCME (red curve) observed by SOHO/LASCO/C2 and STA-HI-1 at different times. (d) Interaction between the preCME and priCME based on the GCS results. (e) Observation of the type II radio burst by WIND/WAVES, associated with the estimated shock heights and speed.} 
	\label{20131028Event}
\end{figure}

In some cases, the conditions of the magnetic connectivity, priCME speed, flare size, suprathermal backgrounds and CME-CME interaction are favorable for generating a strong SEP enhancement, but the 1 AU in situ observation exhibits a much weaker SEP peak intensity. We discuss a specific event for this phenomenon in this section. The SEP event starts at around 05:30 UT on 2013 October 28 and reaches its peak at around 14:00 UT with $I_p=0.91$ pfu for 15--40 MeV protons (corrected by 1.8 to compare with STEREO measurement) by GOES/EPS at L1 point. This is therefore an SEP-poor event with peak intensity much less than the threshold of 4.3 pfu. The preCME and priCME are identified as CMEs with LASCO/C2 first appearance time at 02:24 and 04:48 UT, respectively. Here we would like to make a correction for the preCME: the CDAW catalog identifies the preCME as a single eruption; however, after carefully checking the observations by SDO/AIA and coronagraphs, we determined that this eruption is a complicated structure, combining two separated CMEs which have close propagation directions and speeds (see Figure \ref{20131028Event}(a) and Table \ref{cmeinfo}). In the following part, only the southern preCME is considered as it is the only one with a clear component within the ecliptic plane where STEREO and L1 spacecraft are located. Both the preCME and priCME are associated with strong flares, which are an X1.0-class one for the preceding eruption and an M5.1 one for the primary eruption, respectively. The related information for this event of the priCME speed ($V_{pri}=1249$ km/s), flare size (M5.1), suprathermal backgrounds (1--2 MeV $I_b=85.76$ pfu in the period of 4 hours before the SEP onset) and CME-CME ($d_n=0.98$) information are shown by the cross symbol in Figure \ref{priCME_SEP} to \ref{inte_ip}. This highlights that the conditions of these factors may be favorable for a stronger SEP enhancement, but the estimated $I_0 \sim 0.91$ pfu is much lower than the expectation. Here, we hypothesize that the reason is the blocking by the preceding eruption, and this block may be related to the disturbances of the coronal magnetic fields by the preCME or the trapping of particles inside the preCME.

Figure \ref{20131028Event}(b) shows the priCME observed by LASCO/C2, in which it is found that the CME shape is deformed, i.e., the southern part appears to be contained backward by the preCME. Figure \ref{20131028Event}(c) shows the observations of the preCME (blue front) and the priCME (red front) by Heliospheric Imager‐1 (HI-1) on board STA, which extends the observations from the outer part of COR2-FOV to about a third of the distance to Earth's orbit. Figure \ref{20131028Event}(d) shows the interaction between the preCME and priCME based on the GCS results. The black and green diamonds depict the GCS model heights of the leading edge and trailing edge of the preCME, and red diamonds depict the heights of the leading edge of the priCME. The linear-fit results are given in the figure by the dashed lines, and the orange one is the projection of the red line in the direction of the preCME. The vertical dashed line indicates the particle release time by assuming that particles are released when the priCME reaches the height of 4 Rs. It is found that the particle release occurs when both CMEs start to have meaningful interaction. We expect the direct interaction to occur for at least 2 hours when the priCME propagates from $\sim 4$ to $\sim 16$ Rs but probably even longer and to larger distances in the direction of the preCME. 

The type II radio burst observed by the WIND/WAVES in Figure \ref{20131028Event}(e) indicates the existence of the shock driven by the priCME \citep{2005JGRA..11012S07G,2015ApJ...806...13M,2019SoPh..294..134K}. The profile of the type II radio burst is plotted by the yellow curve, shifting from high-frequencies to low-frequencies, indicating the shock wave propagating from high to low electron density region. This panel also shows the signature of the enhancement of type II radio burst at around 06:00 UT, which confirms the occurrence of the direct CME-CME interaction at this time, and is consistent with the result obtained by the GCS model. To estimate the shock height via the frequency-density relationship, the hybrid model for the background electron density proposed by \citet{2004A&A...413..753V} is used. The red diamonds plot the estimated shock heights, and the dashed line is the linear-fit result, giving the shock speed is $\sim 1326$ km/s. One can refer to \citet{2007SoPh..244..167P} for the details about using the radio observations to estimate the shock propagation characteristics. It is found that the shock height is slightly lower than the height of the priCME front at the same time, but the shock speed is slightly higher. The underestimation of the heights may be due to that (1) the 3-D heights are not considered, (2) the type II radio burst may be generated near the shock flank, or (3) the shock wave propagates inside the preCME and thus the density model is not suitable. However, the relatively decent comparison between the height and speed from GCS and from the type II radio burst confirms the analysis and the fact that there is direct CME-CME interaction. 

Overall, Figure \ref{20131028Event} indicates that during the interaction, the shock and the priCME front interact with the preceding magnetic ejecta (whose propagation direction is nearly in the ecliptic plane) for nearly two hours when the particles are released, which leads to the potential conditions that: (1) the released particles are trapped inside the preCME or (2) the ability of the shock accelerating particles is weakened when the shock is propagating inside the magnetic ejecta with stronger magnetic fields. The effect of a preceding magnetic ejecta on the particle transports was studied by \citet{2001ICRC....8.3273K} via a numerical model. The simulations found that the ejecta would act as an effective barrier for the particle transport if it is ahead of the particle source. We use the velocity dispersion analysis (VDA) to estimate the particle release time and the transport path length. This technique assumes that SEPs in different energies are released at the same time and transport along the same path to the observer. One can refer to \citet{2020ApJ...901...45Z} for the details about using VDA. 
We apply VDA to the ERNE/HED data because HED has more energy channels than GOES/EPS. The results show that the particles are released at 05:25($\pm 18$) UT (the light traveling time of 8.3 minutes is added), and transport along the length of 3.03($\pm 0.47$) AU. The release time derived by VDA is $\sim 35$ minutes later than the result in Figure \ref{20131028Event}(c), and the path length is found to be much higher than the nominal Parker spiral length, which is consistent with the preCME significantly affecting the particle release and/or transport.

\subsection{Influences of the Twin-CME Eruption on SEP Production: Looking at SEP-rich and SEP-poor Events}
\begin{figure}[!hbt]
 	\includegraphics[width=\textwidth]{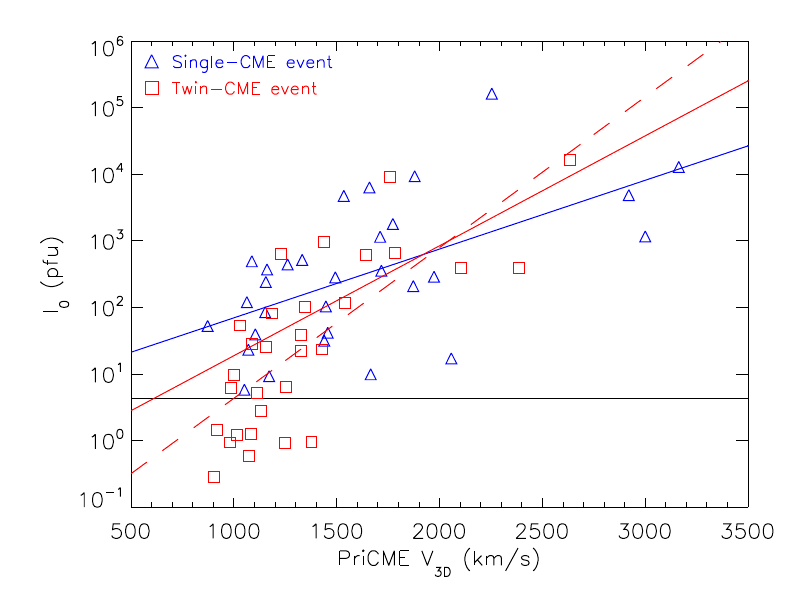}
 	\caption{\small Plot of $V_{pri}$ vs. $I_0$ for the single-(blue) and twin-(red) CME events. The blue line indicates the linear-fit to the single-CME SEP-rich events, and the red dashed and solid lines indicate the linear-fit to the SEP-rich adding SEP-poor events and SEP-rich events, respectively. The horizontal solid line marks the threshold of SEP-rich events.} 
 	\label{singlevstwin}
 \end{figure}

 In this section, we discuss the role of CME-CME interaction in SEP generation. As discussed in Section \ref{sec_multi_role}, there is no correlation between the preCMEs and the seed populations (for our samples). However, it is still worthwhile to study the $V_{pri}$-$I_0$ behaviors for single- or twin- CME events as particles are believed to be more efficiently accelerated in the twin-CME scenario. To do so, we collect the single-CME SEP-rich events (30 events) as given by \citet{2020ApJ...901...45Z}, and estimate the related $I_0$ in 15--40 MeV as done in Figure \ref{int_lon} (the results are not provided here). Figure \ref{singlevstwin} shows the plot of $V_{pri}$ vs. $I_0$ for the single- and twin- CME events.
A linear fit is applied to the single-CME SEP-rich events (blue solid line), twin-CME SEP-rich events (red solid line) and twin-CME SEP-rich adding SEP-poor events (red dashed line), respectively. It is found that the slopes of the red solid line of $0.0017 \pm 0.0003 \ (1\sigma)$ and dashed line of $0.0023 \pm 0.0004 \ (1\sigma)$ are higher than the that of the blue line of  $0.0010 \pm 0.0003 \ (1\sigma)$, providing a direct indication that particle acceleration during twin-CME events is more efficient. Moreover, if only focusing on the SEP-rich events, the percentage of the number of the events with $V_{pri}>1500$ km/s is: $\sim 47\%$ (14/30) for the single-CME events vs. $\sim 33\%$ (7/21) for the twin-CME events. It further indicates that the CME speed in the twin-CME scenario is not necessary to be as high as that associated single-CME eruption to generate large SEP events. However, based on these samples in SC24, the SEP peak intensities in the twin-CME events are not systematically higher than those in the single-CME events, which is different from previous statistical results for the events in SC23 \citep[e.g.,][]{2012AIPC.1436..247G,2013ApJ...763...30D}, although the difference could be due to that we restrict ourselves to twin CMEs with clear spatial overlap based on the GCS reconstruction whereas work on SC23 CMEs rely either only on position angle \citep{2013ApJ...763...30D} or on flare location \citep{2004JGRA..10912105G} to identify spatial overlap.

Furthermore, as discussed in Section \ref{sec_20131028} that the preCME ahead can act as a barrier for the particle transport driven by the priCME, and thus here we discuss the related effects further.
 \begin{figure}[!hbt]
	\includegraphics[width=\textwidth]{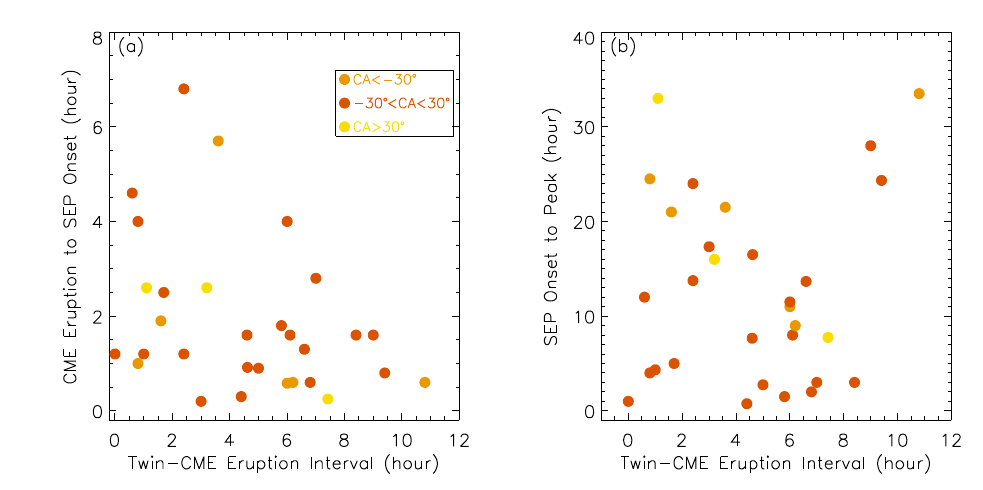}
	\caption{\small (a) Plot of the time interval between twin CMEs vs. the interval between the CME eruption time and the SEP onset time. The related connection angle ($CA$) is shown by different colors. (b) Plot of the time interval between the twin CMEs vs. the interval between SEP onset time and peak time.} 
	\label{pre_to_tr}
\end{figure}
Figure \ref{pre_to_tr}(a) plots the intervals between twin CMEs vs. the time intervals between the CME eruption and the SEP onset, and (b) plots the intervals between twin CMEs vs. the time intervals between SEP onset and peak. Here the LASCO/C2 first appearance time is used as the CME eruption time. It is found that there exist a weak relationship that the larger CME-CME interval resulting in smaller time interval between CME eruption and SEP onset, but the distribution in (b) is scattered. Except for the barrier effect of preCMEs, the relationship between CME interval and this CME-SEP time interval may also be due to that the magnetic connectivity is modified by the eruption of preCME(s) \citep{2009IAUS..257..391L,2010AIPC.1216..440L}. As for considering the SEP onset time to the peak time, the SEP transport effect shall be included.

The results in Figure \ref{inte_ip}(b) and \ref{singlevstwin} support the positive role of twin-CME eruption in the more efficient particle acceleration, but the PCs with $d_n$ combined do not show significant increases as shown in Figure \ref{projection_c}. Here we would like to discuss the two-sided role of the CME-CME interaction. On the positive side in the twin-CME scenario, the eruption of the preCME can provide an enhanced seed populations and turbulence level for the priCME, and the preCME sweep away a part of the coronal plasma and let the shock of the priCME propagate in a region with lower plasma density. On the negative side, the CME-CME interaction may lead to the decrease in the priCME speed \citep{2017SoPh..292...64L}, and the preCME may act as a barrier for the SEP transport if the priCME (and its shock) propagates inside the preceding magnetic ejecta with stronger magnetic field (the case in Section \ref{sec_20131028}). Furthermore, as for the shock propagating inside a preceding magnetic ejecta, \citet{2020ApJ...905....8L} developed an analytical treatment for this scenario. Based on the structure of shock inside magnetic clouds observed at 1 AU, they found that the time for the shock accelerating particles from a lower energy to a higher one is always insufficient. We are looking forward to the Parker Solar Probe approaching the Sun in the future, while the role of the suprathermal particles can be better studied by the observations closer to the Sun \citep[e.g.,][]{2020ApJS..246...33S}.

\section{Summary}\label{summary}
In this paper, we focus on the twin-CME SEP events. We first identify 19 twin-CME events whose SEP enhancements are not strong ($<10$ pfu) as observed by the spacecraft at L1 point (called SEP-poor events). We then combine the observations from the twin STEREO spacecraft, and the fitting equation provided by \citet{2019JGRA..124.6384X} to estimate the peak intensity with the best magnetic connectivity ($I_0$). We find that, among this 19 events, six events can be directly identified as SEP-rich event after using STEREO observations (meaning they are clear SEP-rich events and observed as such but not at L1), four can be identified as SEP-rich events only based on the estimated $I_0$ (meaning they are likely to be SEP-rich but not observed as such by any spacecraft), and nine are SEP-poor events as their $I_0$ values do not exceed the related threshold. This result indicate that the magnetic connectivity in longitude acts as one of the dominant roles in the SEP detection at the spacecraft at 1 AU. Combined with the results of \citet{2020ApJ...901...45Z}, it means that 21 out of 30 twin-CME events in SC24 are SEP-rich as indicated by the fitted peak intensity, $I_0$.

We analyze the role of the priCME speed ($V_{pri}$), the latitude of the propagation direction of the priCME with respect to the ecliptic plane ($\triangle \theta$), the flare size ($I_{SXR}$), the suprathermal backgrounds ($I_b$), and the normalized distance ($d_n$) of CME-CME interaction in the observed and estimated SEP peak intensities based on one-on-one correlations. We find:

(1) Among these independent variables, $V_{pri}$ has the highest CC level with SEP peak intensity. This indicates that the dominant role in determining the SEP peak intensity is the priCME speed. The combination of both SEP-rich and SEP-poor events may help with a better understanding about how CME speed relates to the SEP peak intensity.

(2) The considering of $\triangle \theta$ can lead to the significant increase in CCs for the CME-SEP correlation for the SEP-poor events. It demonstrates the important role of the CME propagation direction in latitude, though the exact form of latitudinal contribution is still not clear.

(3) $I_{SXR}$ is found to have weak correlation with $I_0$ with ${\rm{CC}}=0.384$, while the SEP events with higher $I_0$ tend to be associated with stronger flares. Flare contribution is hard to investigate due to the correlation between CME speed and flare intensity.

(4) $I_b$ is found to correlate with $I_0$ with moderate ${\rm{CC}} \sim 0.5$ for the SEP-rich events, but not for the SEP-poor events. This may indicate suprathermal ion backgrounds act as seed populations for the particle acceleration by CME-driven shock. 

(5) In CME-CME interaction, $d_n$ is found to correlate with $I_0$ only for the SEP-rich events with ${\rm{CC}}= 0.568$. It shows that the SEP peak intensity is higher when the priCME is closer to the preCME when the particles are accelerated and released, which is consistent with the result in \citet{2020ApJ...901...45Z}.

The above results indicate that the priCMEs (e.g., by the factor of $V_{pri}$) act as the dominant role in determining the SEP peak intensities. We analyze the role of the combination of all the parameters by the projection correlation method. It is found that different combinations of these parameters except considering $\triangle \theta$ with $V_{pri}$ do not increase the related projection correlation coefficients, when compared to the result of purely $V_{pri}$ vs. $I_0$. This further support the dominant role of priCMEs in SEP events, and the considering of the latitudinal effect of priCMEs. The results that the combinations do not increase the coefficients may be due to that (1) these factors are closely coupled with priCMEs (such as the flare-CME coupling), or (2) the non-CME roles are more complicated than we can investigate in this paper. In the future, projection correlation method can be applied to more samples with single-CME events and the events in SC23 considered to test the multiple roles in the SEP enhancements.

Finally, the major roles of the magnetic connectivity in longitude as well as latitude and the relatively lower priCME speed, and the potential effect of the preCME may explain the occurrences of the twin-CME events without strong SEP observations at L1 point. Besides, by combining the SC24 data, the twin-CMEs with a fast and wide priCME are found to have higher peak SEP fluxes than the fast and wide single-CMEs, though this needs to be confirmed with a larger sample.

\textbf{Acknowledgement} We acknowledge the use of the data from GOES, SOHO, STEREO, ACE, WIND and SDO. The CME and SEP catalog used in this paper are generated and maintained at the CDAW Data Center by NASA and the Catholic University of America in cooperation with the Naval Research Laboratory. We acknowledge Dr. Wanjun Liu for analyzing the projection correlations. Research for this work was made possible by NASA grants 80NSSC17K0009 and 80NSSC19K0831, and NSF grant AGS1435785. 

\appendix
\section{Estimation of the Best-connected Peak Intensity}\label{app}
Since not all the events have SEP observations from multiple spacecraft, the established fitting equation with given parameters is necessary for estimating the SEP peak intensity with the best magnetic connectivity. In this work, we use the fitting equation with an east-west asymmetry provided by \citet{2019JGRA..124.6384X} (called EqI) instead of the symmetric Gaussian distribution \citep[e.g.,][]{2013ApJ...767...41L,2014SoPh..289.3059R}, which may result in the different estimation of $I_0$. For the purpose of the comparison, we apply the equation provided by \citet{2013ApJ...767...41L} (called EqII) to the same events. This equation is re-written as $I(\phi)=I_0 \exp (- (\phi-\phi_0)^{2} / 2 \sigma^{2})$, where $\phi$ is CA, $\phi_0=-12^\circ+\pm 3^\circ$ and $\sigma=43^\circ \pm 2^\circ$ for 15--40 MeV protons. The fixed values of $\phi_0=-12^\circ$ and $\sigma=43^\circ$ are used for all the events.
\begin{figure}[!hbt]
	\includegraphics[width=\textwidth]{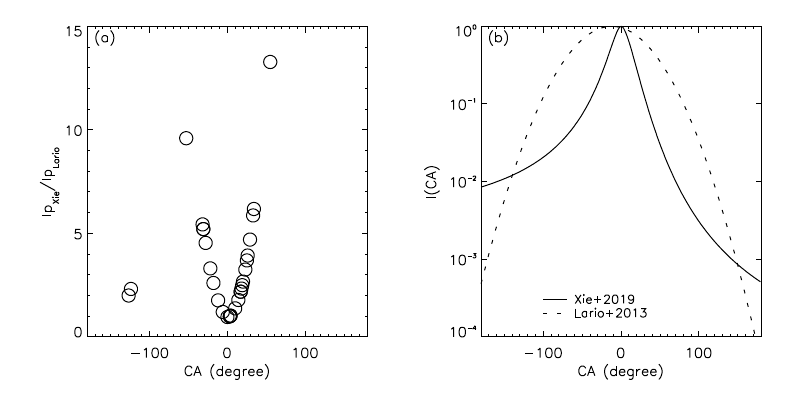}
	\caption{\small (a) Comparison between the $I_0$ estimated by the prediction equations in \citet{2019JGRA..124.6384X} and \citet{2013ApJ...767...41L}. (b) The theoretical profiles of $I(CA)$ with the same $I_0$ estimated by both equations.} 
	\label{xievslario}
\end{figure}
Figure \ref{xievslario}(a) shows the comparison between the $I_0$ estimated by the two prediction equations. It is found that for most of the events, the $I_0$ estimated by EqI is higher than that estimated by EqII, and the difference becomes greater with higher $|CA|$ value when $-65^\circ<CA<70^\circ$. This can be confirmed by comparing the theoretical profiles of $I(CA)$ in Figure \ref{xievslario}(b). In this figure, the highest peak intensity ($I_0$) is set to be unity for both cases. We do not make any initial assumptions regarding which method would be more suitable. We calculate the root-mean-square (RMS) error between the observations and the estimations, and only the events with observations from at lease two spacecraft are used. The RMS error is calculated by Equation \ref{chi}, 
\begin{equation}\label{chi}
			\chi=\sqrt{\frac{\sum_{n=1}^N \sum_{s=1}^3 ( \frac{I_{n,s}-\hat{I}_{n,s}}{I_{n,s}} )}{N}}
\end{equation}
where $n$ and $s$ represent the event and spacecraft, and $I_{n,s}$ and $\hat{I}_{n,s}$ are the observed and estimated $I_p$, respectively. It is found that $\chi$ with EqI is smaller than that with EqII. Moreover, \citet{2015ApJS..218...17H} provided a numerical study of the longitudinal distribution of SEPs. In their Figure 3,  for 20-MeV protons, the longitudinal distribution exhibits an east-west asymmetry while the western profile declines faster than the eastern one, and the decline speed is similar to the decline speed estimated by EqI. Therefore, the usage of EqI in this paper is effective for the estimation of $I_0$.


\begin{thebibliography}{}
	\expandafter\ifx\csname natexlab\endcsname\relax\def\natexlab#1{#1}\fi
	
	\bibitem[{{Brueckner} {et~al.}(1995){Brueckner}, {Howard}, {Koomen},
		{Korendyke}, {Michels}, {Moses}, {Socker}, {Dere}, {Lamy}, {Llebaria},
		{Bout}, {Schwenn}, {Simnett}, {Bedford}, \& {Eyles}}]{1995SoPh..162..357B}
	{Brueckner}, G.~E., {Howard}, R.~A., {Koomen}, M.~J., {et~al.} 1995, \solphys,
	162, 357
	
	\bibitem[{{Cane} {et~al.}(2010){Cane}, {Richardson}, \& {von
			Rosenvinge}}]{2010JGRA..115.8101C}
	{Cane}, H.~V., {Richardson}, I.~G., \& {von Rosenvinge}, T.~T. 2010, Journal of
	Geophysical Research (Space Physics), 115, A08101
	
	\bibitem[{{Cheng} {et~al.}(2010){Cheng}, {Zhang}, {Ding}, \&
		{Poomvises}}]{2010ApJ...712..752C}
	{Cheng}, X., {Zhang}, J., {Ding}, M.~D., \& {Poomvises}, W. 2010, \apj, 712,
	752
	
	\bibitem[{{Dalla} \& {Agueda}(2010)}]{2010AIPC.1216..613D}
	{Dalla}, S., \& {Agueda}, N. 2010, in American Institute of Physics Conference
	Series, Vol. 1216, Twelfth International Solar Wind Conference, ed.
	M.~{Maksimovic}, K.~{Issautier}, N.~{Meyer-Vernet}, M.~{Moncuquet}, \&
	F.~{Pantellini}, 613--616
	
	\bibitem[{{Desai} \& {Giacalone}(2016)}]{2016LRSP...13....3D}
	{Desai}, M., \& {Giacalone}, J. 2016, Living Reviews in Solar Physics, 13, 3
	
	\bibitem[{{Desai} {et~al.}(2006){Desai}, {Mason}, {Gold}, {Krimigis}, {Cohen},
		{Mewaldt}, {Mazur}, \& {Dwyer}}]{2006ApJ...649..470D}
	{Desai}, M.~I., {Mason}, G.~M., {Gold}, R.~E., {et~al.} 2006, \apj, 649, 470
	
	\bibitem[{{Dierckxsens} {et~al.}(2015){Dierckxsens}, {Tziotziou}, {Dalla},
		{Patsou}, {Marsh}, {Crosby}, {Malandraki}, \&
		{Tsiropoula}}]{2015SoPh..290..841D}
	{Dierckxsens}, M., {Tziotziou}, K., {Dalla}, S., {et~al.} 2015, \solphys, 290,
	841
	
	\bibitem[{{Ding} {et~al.}(2013){Ding}, {Jiang}, {Zhao}, \&
		{Li}}]{2013ApJ...763...30D}
	{Ding}, L., {Jiang}, Y., {Zhao}, L., \& {Li}, G. 2013, \apj, 763, 30
	
	\bibitem[{{Ding} {et~al.}(2015){Ding}, {Li}, {Le}, {Gu}, \&
		{Cao}}]{2015ApJ...812..171D}
	{Ding}, L.-G., {Li}, G., {Le}, G.-M., {Gu}, B., \& {Cao}, X.-X. 2015, \apj,
	812, 171
	
	\bibitem[{{Ding} {et~al.}(2019){Ding}, {Wang}, {Feng}, {Li}, \&
		{Jiang}}]{2019RAA....19....5D}
	{Ding}, L.-G., {Wang}, Z.-W., {Feng}, L., {Li}, G., \& {Jiang}, Y. 2019,
	Research in Astronomy and Astrophysics, 19, 005
	
	\bibitem[{{Ding} {et~al.}(2014){Ding}, {Li}, {Jiang}, {Le}, {Shen}, {Wang},
		{Chen}, {Xu}, {Gu}, \& {Zhang}}]{2014ApJ...793L..35D}
	{Ding}, L.-G., {Li}, G., {Jiang}, Y., {et~al.} 2014, \apjl, 793, L35
	
	\bibitem[{{Domingo} {et~al.}(1995){Domingo}, {Fleck}, \& {Poland}}]{domingo95}
	{Domingo}, V., {Fleck}, B., \& {Poland}, A.~I. 1995, 162, 1
	
	\bibitem[{{Dresing} {et~al.}(2014){Dresing}, {G{\'o}mez-Herrero}, {Heber},
		{Klassen}, {Malandraki}, {Dr{\"o}ge}, \& {Kartavykh}}]{2014A&A...567A..27D}
	{Dresing}, N., {G{\'o}mez-Herrero}, R., {Heber}, B., {et~al.} 2014, \aap, 567,
	A27
	
	\bibitem[{{Giacalone} \& {Jokipii}(2012)}]{2012AIPC.1436..130G}
	{Giacalone}, J., \& {Jokipii}, J.~R. 2012, in American Institute of Physics
	Conference Series, Vol. 1436, Physics of the Heliosphere: A 10 Year
	Retrospective, ed. J.~{Heerikhuisen}, G.~{Li}, N.~{Pogorelov}, \& G.~{Zank},
	130--135
	
	\bibitem[{{Gold} {et~al.}(1998){Gold}, {Krimigis}, {Hawkins}, {Haggerty},
		{Lohr}, {Fiore}, {Armstrong}, {Holland}, \&
		{Lanzerotti}}]{1998SSRv...86..541G}
	{Gold}, R.~E., {Krimigis}, S.~M., {Hawkins}, S.~E., I., {et~al.} 1998, \ssr,
	86, 541
	
	\bibitem[{{Gopalswamy}(2012)}]{2012AIPC.1436..247G}
	{Gopalswamy}, N. 2012, in American Institute of Physics Conference Series, Vol.
	1436, Physics of the Heliosphere: A 10 Year Retrospective, ed.
	J.~{Heerikhuisen}, G.~{Li}, N.~{Pogorelov}, \& G.~{Zank}, 247--252
	
	\bibitem[{{Gopalswamy} {et~al.}(2005){Gopalswamy}, {Aguilar-Rodriguez},
		{Yashiro}, {Nunes}, {Kaiser}, \& {Howard}}]{2005JGRA..11012S07G}
	{Gopalswamy}, N., {Aguilar-Rodriguez}, E., {Yashiro}, S., {et~al.} 2005,
	Journal of Geophysical Research (Space Physics), 110, A12S07
	
	\bibitem[{{Gopalswamy} {et~al.}(2014){Gopalswamy}, {Xie}, {Akiyama},
		{M{\"a}kel{\"a}}, \& {Yashiro}}]{2014EP&S...66..104G}
	{Gopalswamy}, N., {Xie}, H., {Akiyama}, S., {M{\"a}kel{\"a}}, P.~A., \&
	{Yashiro}, S. 2014, Earth, Planets, and Space, 66, 104
	
	\bibitem[{{Gopalswamy} {et~al.}(2001){Gopalswamy}, {Yashiro}, {Kaiser},
		{Howard}, \& {Bougeret}}]{2001ApJ...548L..91G}
	{Gopalswamy}, N., {Yashiro}, S., {Kaiser}, M.~L., {Howard}, R.~A., \&
	{Bougeret}, J.~L. 2001, \apjl, 548, L91
	
	\bibitem[{{Gopalswamy} {et~al.}(2004){Gopalswamy}, {Yashiro}, {Krucker},
		{Stenborg}, \& {Howard}}]{2004JGRA..10912105G}
	{Gopalswamy}, N., {Yashiro}, S., {Krucker}, S., {Stenborg}, G., \& {Howard},
	R.~A. 2004, Journal of Geophysical Research (Space Physics), 109, A12105
	
	\bibitem[{{Gopalswamy} {et~al.}(2003){Gopalswamy}, {Yashiro}, {Lara}, {Kaiser},
		{Thompson}, {Gallagher}, \& {Howard}}]{2003GeoRL..30.8015G}
	{Gopalswamy}, N., {Yashiro}, S., {Lara}, A., {et~al.} 2003, \grl, 30, 8015
	
	\bibitem[{{Gopalswamy} {et~al.}(2002){Gopalswamy}, {Yashiro}, {Micha{\l}ek},
		{Kaiser}, {Howard}, {Reames}, {Leske}, \& {von
			Rosenvinge}}]{2002ApJ...572L.103G}
	{Gopalswamy}, N., {Yashiro}, S., {Micha{\l}ek}, G., {et~al.} 2002, \apjl, 572,
	L103
	
	\bibitem[{{Gopalswamy} {et~al.}(2009){Gopalswamy}, {Yashiro}, {Michalek},
		{Stenborg}, {Vourlidas}, {Freeland}, \& {Howard}}]{2009EM&P..104..295G}
	{Gopalswamy}, N., {Yashiro}, S., {Michalek}, G., {et~al.} 2009, Earth Moon and
	Planets, 104, 295
	
	\bibitem[{{Gou} {et~al.}(2020){Gou}, {Veronig}, {Liu}, {Zhuang},
		{Dumbovi{\'c}}, {Podladchikova}, {Reid}, {Temmer}, {Dissauer},
		{Vr{\v{s}}nak}, \& {Wang}}]{2020ApJ...897L..36G}
	{Gou}, T., {Veronig}, A.~M., {Liu}, R., {et~al.} 2020, \apjl, 897, L36
	
	\bibitem[{{He} \& {Wan}(2015)}]{2015ApJS..218...17H}
	{He}, H.~Q., \& {Wan}, W. 2015, \apjs, 218, 17
	
	\bibitem[{{Howard} {et~al.}(2008){Howard}, {Moses}, {Vourlidas}, {Newmark},
		{Socker}, {Plunkett}, {Korendyke}, {Cook}, {Hurley}, {Davila}, {Thompson},
		{St Cyr}, {Mentzell}, {Mehalick}, {Lemen}, {Wuelser}, {Duncan}, {Tarbell},
		{Wolfson}, {Moore}, {Harrison}, {Waltham}, {Lang}, {Davis}, {Eyles},
		{Mapson-Menard}, {Simnett}, {Halain}, {Defise}, {Mazy}, {Rochus}, {Mercier},
		{Ravet}, {Delmotte}, {Auchere}, {Delaboudiniere}, {Bothmer}, {Deutsch},
		{Wang}, {Rich}, {Cooper}, {Stephens}, {Maahs}, {Baugh}, {McMullin}, \&
		{Carter}}]{2008SSRv..136...67H}
	{Howard}, R.~A., {Moses}, J.~D., {Vourlidas}, A., {et~al.} 2008, \ssr, 136, 67
	
	\bibitem[{{Kahler} \& {Ling}(2019)}]{2019ApJ...872...89K}
	{Kahler}, S.~W., \& {Ling}, A.~G. 2019, \apj, 872, 89
	
	\bibitem[{{Kahler} {et~al.}(2019){Kahler}, {Ling}, \&
		{Gopalswamy}}]{2019SoPh..294..134K}
	{Kahler}, S.~W., {Ling}, A.~G., \& {Gopalswamy}, N. 2019, \solphys, 294, 134
	
	\bibitem[{{Kahler} \& {Vourlidas}(2005)}]{2005JGRA..11012S01K}
	{Kahler}, S.~W., \& {Vourlidas}, A. 2005, Journal of Geophysical Research
	(Space Physics), 110, A12S01
	
	\bibitem[{{Kahler} \& {Vourlidas}(2014{\natexlab{a}})}]{2014ApJ...784...47K}
	---. 2014{\natexlab{a}}, \apj, 784, 47
	
	\bibitem[{{Kahler} \& {Vourlidas}(2014{\natexlab{b}})}]{2014ApJ...791....4K}
	---. 2014{\natexlab{b}}, \apj, 791, 4
	
	\bibitem[{{Kaiser} {et~al.}(2008){Kaiser}, {Kucera}, {Davila}, {St. Cyr},
		{Guhathakurta}, \& {Christian}}]{2008SSRv..136....5K}
	{Kaiser}, M.~L., {Kucera}, T.~A., {Davila}, J.~M., {et~al.} 2008, \ssr, 136, 5
	
	\bibitem[{{Kallenrode}(1997)}]{1997JGR...10222335K}
	{Kallenrode}, M.-B. 1997, \jgr, 102, 22335
	
	\bibitem[{{Kallenrode}(2001)}]{2001ICRC....8.3273K}
	{Kallenrode}, M.~B. 2001, in International Cosmic Ray Conference, Vol.~8,
	International Cosmic Ray Conference, 3273
	
	\bibitem[{{Kihara} {et~al.}(2020){Kihara}, {Huang}, {Nishimura}, {Nitta},
		{Yashiro}, {Ichimoto}, \& {Asai}}]{2020ApJ...900...75K}
	{Kihara}, K., {Huang}, Y., {Nishimura}, N., {et~al.} 2020, \apj, 900, 75
	
	\bibitem[{{Kouloumvakos} {et~al.}(2016){Kouloumvakos}, {Patsourakos}, {Nindos},
		{Vourlidas}, {Anastasiadis}, {Hillaris}, \& {Sandberg}}]{2016ApJ...821...31K}
	{Kouloumvakos}, A., {Patsourakos}, S., {Nindos}, A., {et~al.} 2016, \apj, 821,
	31
	
	\bibitem[{{Kouloumvakos} {et~al.}(2019){Kouloumvakos}, {Rouillard}, {Wu},
		{Vainio}, {Vourlidas}, {Plotnikov}, {Afanasiev}, \&
		{{\"O}nel}}]{2019ApJ...876...80K}
	{Kouloumvakos}, A., {Rouillard}, A.~P., {Wu}, Y., {et~al.} 2019, \apj, 876, 80
	
	\bibitem[{{Kozarev} {et~al.}(2015){Kozarev}, {Raymond}, {Lobzin}, \&
		{Hammer}}]{2015ApJ...799..167K}
	{Kozarev}, K.~A., {Raymond}, J.~C., {Lobzin}, V.~V., \& {Hammer}, M. 2015,
	\apj, 799, 167
	
	\bibitem[{{Lario} {et~al.}(2013){Lario}, {Aran}, {G{\'o}mez-Herrero},
		{Dresing}, {Heber}, {Ho}, {Decker}, \& {Roelof}}]{2013ApJ...767...41L}
	{Lario}, D., {Aran}, A., {G{\'o}mez-Herrero}, R., {et~al.} 2013, \apj, 767, 41
	
	\bibitem[{{Lario} {et~al.}(2016){Lario}, {Kwon}, {Vourlidas}, {Raouafi},
		{Haggerty}, {Ho}, {Anderson}, {Papaioannou}, {G{\'o}mez-Herrero}, {Dresing},
		\& {Riley}}]{2016ApJ...819...72L}
	{Lario}, D., {Kwon}, R.~Y., {Vourlidas}, A., {et~al.} 2016, \apj, 819, 72
	
	\bibitem[{{Lario} {et~al.}(2020){Lario}, {Kwon}, {Balmaceda}, {Richardson},
		{Krupar}, {Thompson}, {Cyr}, {Zhao}, \& {Zhang}}]{2020ApJ...889...92L}
	{Lario}, D., {Kwon}, R.~Y., {Balmaceda}, L., {et~al.} 2020, \apj, 889, 92
	
	\bibitem[{{Lee}(1983)}]{1983JGR....88.6109L}
	{Lee}, M.~A. 1983, \jgr, 88, 6109
	
	\bibitem[{{Lemen} {et~al.}(2012){Lemen}, {Title}, {Akin}, {Boerner}, {Chou},
		{Drake}, {Duncan}, {Edwards}, {Friedlaender}, {Heyman}, {Hurlburt}, {Katz},
		{Kushner}, {Levay}, {Lindgren}, {Mathur}, {McFeaters}, {Mitchell}, {Rehse},
		{Schrijver}, {Springer}, {Stern}, {Tarbell}, {Wuelser}, {Wolfson}, {Yanari},
		{Bookbinder}, {Cheimets}, {Caldwell}, {Deluca}, {Gates}, {Golub}, {Park},
		{Podgorski}, {Bush}, {Scherrer}, {Gummin}, {Smith}, {Auker}, {Jerram},
		{Pool}, {Soufli}, {Windt}, {Beardsley}, {Clapp}, {Lang}, \&
		{Waltham}}]{2012SoPh..275...17L}
	{Lemen}, J.~R., {Title}, A.~M., {Akin}, D.~J., {et~al.} 2012, \solphys, 275, 17
	
	\bibitem[{{Li} \& {Lugaz}(2020)}]{2020ApJ...905....8L}
	{Li}, G., \& {Lugaz}, N. 2020, \apj, 905, 8
	
	\bibitem[{{Li} {et~al.}(2012){Li}, {Moore}, {Mewaldt}, {Zhao}, \&
		{Labrador}}]{2012SSRv..171..141L}
	{Li}, G., {Moore}, R., {Mewaldt}, R.~A., {Zhao}, L., \& {Labrador}, A.~W. 2012,
	\ssr, 171, 141
	
	\bibitem[{Liu {et~al.}(2020)Liu, Ke, Liu, \& Li}]{liu2020model}
	Liu, W., Ke, Y., Liu, J., \& Li, R. 2020, Journal of the American Statistical
	Association, 1
	
	\bibitem[{{Lugaz} {et~al.}(2009){Lugaz}, {Roussev}, \&
		{Sokolov}}]{2009IAUS..257..391L}
	{Lugaz}, N., {Roussev}, I.~I., \& {Sokolov}, I.~V. 2009, in Universal
	Heliophysical Processes, ed. N.~{Gopalswamy} \& D.~F. {Webb}, Vol. 257,
	391--398
	
	\bibitem[{{Lugaz} {et~al.}(2010){Lugaz}, {Roussev}, {Sokolov}, \&
		{Jacobs}}]{2010AIPC.1216..440L}
	{Lugaz}, N., {Roussev}, I.~I., {Sokolov}, I.~V., \& {Jacobs}, C. 2010, in
	American Institute of Physics Conference Series, Vol. 1216, Twelfth
	International Solar Wind Conference, ed. M.~{Maksimovic}, K.~{Issautier},
	N.~{Meyer-Vernet}, M.~{Moncuquet}, \& F.~{Pantellini}, 440--443
	
	\bibitem[{{Lugaz} {et~al.}(2017){Lugaz}, {Temmer}, {Wang}, \&
		{Farrugia}}]{2017SoPh..292...64L}
	{Lugaz}, N., {Temmer}, M., {Wang}, Y., \& {Farrugia}, C.~J. 2017, \solphys,
	292, 64
	
	\bibitem[{{M{\"a}kel{\"a}} {et~al.}(2015){M{\"a}kel{\"a}}, {Gopalswamy},
		{Akiyama}, {Xie}, \& {Yashiro}}]{2015ApJ...806...13M}
	{M{\"a}kel{\"a}}, P., {Gopalswamy}, N., {Akiyama}, S., {Xie}, H., \& {Yashiro},
	S. 2015, \apj, 806, 13
	
	\bibitem[{{Mason} {et~al.}(2005){Mason}, {Desai}, {Mazur}, \&
		{Dwyer}}]{2005AIPC..781..219M}
	{Mason}, G.~M., {Desai}, M.~I., {Mazur}, J.~E., \& {Dwyer}, J.~R. 2005, in
	American Institute of Physics Conference Series, Vol. 781, The Physics of
	Collisionless Shocks: 4th Annual IGPP International Astrophysics Conference,
	ed. G.~{Li}, G.~P. {Zank}, \& C.~T. {Russell}, 219--226
	
	\bibitem[{{Mason} {et~al.}(1999){Mason}, {Mazur}, \&
		{Dwyer}}]{1999ApJ...525L.133M}
	{Mason}, G.~M., {Mazur}, J.~E., \& {Dwyer}, J.~R. 1999, \apjl, 525, L133
	
	\bibitem[{{Masson} {et~al.}(2013){Masson}, {Antiochos}, \&
		{DeVore}}]{2013ApJ...771...82M}
	{Masson}, S., {Antiochos}, S.~K., \& {DeVore}, C.~R. 2013, \apj, 771, 82
	
	\bibitem[{{Masson} {et~al.}(2019){Masson}, {Antiochos}, \&
		{DeVore}}]{2019ApJ...884..143M}
	---. 2019, \apj, 884, 143
	
	\bibitem[{{Mewaldt} {et~al.}(2008){Mewaldt}, {Cohen}, {Cook}, {Cummings},
		{Davis}, {Geier}, {Kecman}, {Klemic}, {Labrador}, {Leske}, {Miyasaka},
		{Nguyen}, {Ogliore}, {Stone}, {Radocinski}, {Wiedenbeck}, {Hawk}, {Shuman},
		{von Rosenvinge}, \& {Wortman}}]{2008SSRv..136..285M}
	{Mewaldt}, R.~A., {Cohen}, C.~M.~S., {Cook}, W.~R., {et~al.} 2008, \ssr, 136,
	285
	
	\bibitem[{{Mewaldt} {et~al.}(2012){Mewaldt}, {Looper}, {Cohen}, {Haggerty},
		{Labrador}, {Leske}, {Mason}, {Mazur}, \& {von
			Rosenvinge}}]{2012SSRv..171...97M}
	{Mewaldt}, R.~A., {Looper}, M.~D., {Cohen}, C.~M.~S., {et~al.} 2012, \ssr, 171,
	97
	
	\bibitem[{{Miteva} {et~al.}(2013){Miteva}, {Klein}, {Malandraki}, \&
		{Dorrian}}]{2013SoPh..282..579M}
	{Miteva}, R., {Klein}, K.~L., {Malandraki}, O., \& {Dorrian}, G. 2013,
	\solphys, 282, 579
	
	\bibitem[{{M{\"u}ller-Mellin} {et~al.}(2008){M{\"u}ller-Mellin},
		{B{\"o}ttcher}, {Falenski}, {Rode}, {Duvet}, {Sanderson}, {Butler},
		{Johlander}, \& {Smit}}]{2008SSRv..136..363M}
	{M{\"u}ller-Mellin}, R., {B{\"o}ttcher}, S., {Falenski}, J., {et~al.} 2008,
	\ssr, 136, 363
	
	\bibitem[{Palmerio {et~al.}(2021)Palmerio, Kilpua, Witasse, Barnes,
		Sánchez-Cano, Weiss, Nieves-Chinchilla, Möstl, Jian, Mierla, Zhukov, Guo,
		Rodriguez, Lowrance, Isavnin, Turc, Futaana, \& Holmström}]{Palmerio_2021}
	Palmerio, E., Kilpua, E. K.~J., Witasse, O., {et~al.} 2021, Space Weather, n/a,
	e2020SW002654, e2020SW002654 2020SW002654
	
	\bibitem[{{Park} \& {Moon}(2014)}]{2014JGRA..119.9456P}
	{Park}, J., \& {Moon}, Y.~J. 2014, Journal of Geophysical Research (Space
	Physics), 119, 9456
	
	\bibitem[{{Pesnell} {et~al.}(2012){Pesnell}, {Thompson}, \&
		{Chamberlin}}]{2012SoPh..275....3P}
	{Pesnell}, W.~D., {Thompson}, B.~J., \& {Chamberlin}, P.~C. 2012, \solphys,
	275, 3
	
	\bibitem[{{Pohjolainen} {et~al.}(2007){Pohjolainen}, {van Driel-Gesztelyi},
		{Culhane}, {Manoharan}, \& {Elliott}}]{2007SoPh..244..167P}
	{Pohjolainen}, S., {van Driel-Gesztelyi}, L., {Culhane}, J.~L., {Manoharan},
	P.~K., \& {Elliott}, H.~A. 2007, \solphys, 244, 167
	
	\bibitem[{{Reames}(1999)}]{1999SSRv...90..413R}
	{Reames}, D.~V. 1999, \ssr, 90, 413
	
	\bibitem[{{Reames}(2013)}]{2013SSRv..175...53R}
	---. 2013, \ssr, 175, 53
	
	\bibitem[{{Richardson} {et~al.}(2014){Richardson}, {von Rosenvinge}, {Cane},
		{Christian}, {Cohen}, {Labrador}, {Leske}, {Mewaldt}, {Wiedenbeck}, \&
		{Stone}}]{2014SoPh..289.3059R}
	{Richardson}, I.~G., {von Rosenvinge}, T.~T., {Cane}, H.~V., {et~al.} 2014,
	\solphys, 289, 3059
	
	\bibitem[{{Rouillard} {et~al.}(2012){Rouillard}, {Sheeley}, {Tylka},
		{Vourlidas}, {Ng}, {Rakowski}, {Cohen}, {Mewaldt}, {Mason}, {Reames},
		{Savani}, {StCyr}, \& {Szabo}}]{2012ApJ...752...44R}
	{Rouillard}, A.~P., {Sheeley}, N.~R., {Tylka}, A., {et~al.} 2012, \apj, 752, 44
	
	\bibitem[{{Schwadron} {et~al.}(2020){Schwadron}, {Bale}, {Bonnell}, {Case},
		{Christian}, {Cohen}, {Cummings}, {Davis}, {Dudok de Wit}, {de Wet}, {Desai},
		{Joyce}, {Goetz}, {Giacalone}, {Gorby}, {Harvey}, {Heber}, {Hill},
		{Karavolos}, {Kasper}, {Korreck}, {Larson}, {Livi}, {Leske}, {Malandraki},
		{MacDowall}, {Malaspina}, {Matthaeus}, {McComas}, {McNutt}, {Mewaldt},
		{Mitchell}, {Mays}, {Niehof}, {Odstrcil}, {Pulupa}, {Poduval}, {Rankin},
		{Roelof}, {Stevens}, {Stone}, {Szalay}, {Wiedenbeck}, {Winslow}, \&
		{Whittlesey}}]{2020ApJS..246...33S}
	{Schwadron}, N.~A., {Bale}, S., {Bonnell}, J., {et~al.} 2020, \apjs, 246, 33
	
	\bibitem[{{Shen} {et~al.}(2013){Shen}, {Li}, {Kong}, {Hu}, {Sun}, {Ding},
		{Chen}, {Wang}, \& {Xia}}]{2013ApJ...763..114S}
	{Shen}, C., {Li}, G., {Kong}, X., {et~al.} 2013, \apj, 763, 114
	
	\bibitem[{{Takahashi} {et~al.}(2016){Takahashi}, {Mizuno}, \&
		{Shibata}}]{2016ApJ...833L...8T}
	{Takahashi}, T., {Mizuno}, Y., \& {Shibata}, K. 2016, \apjl, 833, L8
	
	\bibitem[{{Thernisien} {et~al.}(2009){Thernisien}, {Vourlidas}, \&
		{Howard}}]{2009SoPh..256..111T}
	{Thernisien}, A., {Vourlidas}, A., \& {Howard}, R.~A. 2009, \solphys, 256, 111
	
	\bibitem[{{Thernisien} {et~al.}(2006){Thernisien}, {Howard}, \&
		{Vourlidas}}]{2006ApJ...652..763T}
	{Thernisien}, A.~F.~R., {Howard}, R.~A., \& {Vourlidas}, A. 2006, \apj, 652,
	763
	
	\bibitem[{{Torsti} {et~al.}(1995){Torsti}, {Valtonen}, {Lumme}, {Peltonen},
		{Eronen}, {Louhola}, {Riihonen}, {Schultz}, {Teittinen}, {Ahola}, {Holmlund},
		{Kelh{\"a}}, {Lepp{\"a}l{\"a}}, {Ruuska}, \&
		{Str{\"o}mmer}}]{1995SoPh..162..505T}
	{Torsti}, J., {Valtonen}, E., {Lumme}, M., {et~al.} 1995, \solphys, 162, 505
	
	\bibitem[{{Trottet} {et~al.}(2015){Trottet}, {Samwel}, {Klein}, {Dudok de Wit},
		\& {Miteva}}]{2015SoPh..290..819T}
	{Trottet}, G., {Samwel}, S., {Klein}, K.~L., {Dudok de Wit}, T., \& {Miteva},
	R. 2015, \solphys, 290, 819
	
	\bibitem[{{von Rosenvinge} {et~al.}(2008){von Rosenvinge}, {Reames}, {Baker},
		{Hawk}, {Nolan}, {Ryan}, {Shuman}, {Wortman}, {Mewaldt}, {Cummings}, {Cook},
		{Labrador}, {Leske}, \& {Wiedenbeck}}]{2008SSRv..136..391V}
	{von Rosenvinge}, T.~T., {Reames}, D.~V., {Baker}, R., {et~al.} 2008, \ssr,
	136, 391
	
	\bibitem[{{Vr{\v{s}}nak} {et~al.}(2004){Vr{\v{s}}nak}, {Magdaleni{\'c}}, \&
		{Zlobec}}]{2004A&A...413..753V}
	{Vr{\v{s}}nak}, B., {Magdaleni{\'c}}, J., \& {Zlobec}, P. 2004, \aap, 413, 753
	
	\bibitem[{{Xie} {et~al.}(2019){Xie}, {St. Cyr}, {M{\"a}kel{\"a}}, \&
		{Gopalswamy}}]{2019JGRA..124.6384X}
	{Xie}, H., {St. Cyr}, O.~C., {M{\"a}kel{\"a}}, P., \& {Gopalswamy}, N. 2019,
	Journal of Geophysical Research (Space Physics), 124, 6384
	
	\bibitem[{{Yashiro} \& {Gopalswamy}(2009)}]{2009IAUS..257..233Y}
	{Yashiro}, S., \& {Gopalswamy}, N. 2009, in Universal Heliophysical Processes,
	ed. N.~{Gopalswamy} \& D.~F. {Webb}, Vol. 257, 233--243
	
	\bibitem[{{Yashiro} {et~al.}(2004){Yashiro}, {Gopalswamy}, {Michalek}, {St.
			Cyr}, {Plunkett}, {Rich}, \& {Howard}}]{2004JGRA..109.7105Y}
	{Yashiro}, S., {Gopalswamy}, N., {Michalek}, G., {et~al.} 2004, Journal of
	Geophysical Research (Space Physics), 109, A07105
	
	\bibitem[{{Zhang} {et~al.}(2020){Zhang}, {Temmer}, {Gopalswamy}, {Malandraki},
		{Nitta}, {Patsourakos}, {Shen}, {Vr{\v{s}}nak}, {Wang}, {Webb}, {Desai},
		{Dissauer}, {Dresing}, {Dumbovi{\'c}}, {Feng}, {Heinemann}, {Laurenza},
		{Lugaz}, \& {Zhuang}}]{2020arXiv201206116Z}
	{Zhang}, J., {Temmer}, M., {Gopalswamy}, N., {et~al.} 2020, arXiv e-prints,
	arXiv:2012.06116
	
	\bibitem[{Zhu {et~al.}(2017)Zhu, Xu, Li, \& Zhong}]{zhu2017projection}
	Zhu, L., Xu, K., Li, R., \& Zhong, W. 2017, Biometrika, 104, 829
	
	\bibitem[{{Zhuang} {et~al.}(2020){Zhuang}, {Lugaz}, {Gou}, {Ding}, \&
		{Wang}}]{2020ApJ...901...45Z}
	{Zhuang}, B., {Lugaz}, N., {Gou}, T., {Ding}, L., \& {Wang}, Y. 2020, \apj,
	901, 45
\end{thebibliography}
\end{document}